\documentclass[a4paper,fleqn,usenatbib]{mnras}

\usepackage{newtxtext,newtxmath}

\usepackage[T1]{fontenc}
\usepackage{ae,aecompl}

\usepackage{graphicx}	
\graphicspath{{figures/}} 
\usepackage{amsmath}	
\usepackage{amssymb}	
\usepackage{stmaryrd}
\usepackage{multirow}

\title[Rate of planet-star coalescences]{The Rate of Planet-star Coalescences Due to Tides  and Stellar Evolution}
\author[A. V. Popkov \& S. B. Popov]{
Alexander V. Popkov,$^{1,2}$\thanks{Corresponding author. E-mail: popkov.av@mipt.ru}
Sergei B. Popov$^{3,4}$
\\
$^{1}$Moscow Institute of Physics and Technology, Dolgoprudny, Institutsky~per.~9, Moscow region, 141700, Russia\\
$^{2}$Astro Space Center of Lebedev Physical Institute, Profsoyuznaya~St.~84/32, Moscow, 117997, Russia\\
$^{3}$Sternberg Astronomical Institute, Lomonosov Moscow State University, Universitetsky~pr.~13, Moscow, 119234, Russia\\
$^{4}$National Research University ``Higher School of Economics'', Department of Physics,
Myasnitskaya~str.~20, Moscow, 101000, Russia
}

\date{Accepted XXX. Received YYY; in original form ZZZ}

\pubyear{2019}

\begin{document}
\label{firstpage}
\pagerange{\pageref{firstpage}--\pageref{lastpage}}
\maketitle

\begin{abstract}
Orbits of close-in planets can shrink significantly due to dissipation of tidal energy in a host star. This process can result in star-planet coalescence within the Galactic lifetime. In some cases, such events can be accompanied by an optical or/and UV/X-ray transient. Potentially, these outbursts can be observed in near future with new facilities such as LSST from distances about few Mpc. We use a population synthesis model to study this process and derive the rate of star-planet mergers of different types. Mostly, planets are absorbed by red giants. However, these events, happening with the rate about 3 per year, mostly do not produce detectable transients. The rate of mergers with main sequence stars depends on the effectiveness of tidal dissipation; for reasonable values of stellar tidal quality factor, such events happen in a Milky Way-like galaxy approximately once in 70 yrs or more rarely.
This rate is dominated by planets with low masses. Such events do not produce bright transients having maximum luminosities $\lesssim 10^{36.5}$~erg~s$^{-1}$. Brighter events, related to massive planets, with maximum luminosity $\sim 10^{37.5}$~--~$10^{38}$~erg~s$^{-1}$, have the rate nearly five times smaller.
\end{abstract}




\section{Introduction}

Exoplanetary studies are just about 25
years old. However, already several thousand of planets are known and a huge number of candidates exists.\footnote{See regularly updated on-line catalogues at \url{http://exoplanet.eu/} \citep{2011A&A...532A..79S}, \url{http://exoplanets.org/} \citep{2011PASP..123..412W}, and \url{https://exoplanetarchive.ipac.caltech.edu/} \citep{2013PASP..125..989A}.} Modern observations and their analysis suggest that majority of stars have planetary systems (see e.g. \citealt{Winn15_occurrence, Traub16_occurrence, Foreman16_occurrence, Bryan16_occurrence, 2018haex.bookE.155B, 2018haex.bookE.195W, 2019arXiv190201417H} and references therein).
Many efforts are made to advance our understanding of formation, life, and death of planetary systems.
In the aftermath of turbulent early years, in most of the systems planetary orbits usually become more or less stable for a long period of time. However, secular changes of orbital parameters can go on due to several processes: planet-planet interactions, influence of stellar binary components, near-by passages of stars, etc.

Some of planets, including massive ones, can appear close to their hosts mainly due to migration in protoplanetary discs (see a review in \citealt{2019SAAS...45..151K, Paardekooper2018}, and references therein). Through their lifetime planets with smaller orbits can intensively interact with stars, mostly due to tides \citep{2008EAS....29....1M, Jackson08, Ogilvie14_tidal_review, 2017MNRAS.470.2054C, 2018A&A...618A..18R, 2019A&A...621A.124B}. Often this results in orbit shrinking and sometimes in infall of a planet onto its host star \citep{Teitler14_dearth, Zhang_dizzy, Matsakos15_breakfast, Metzger17}. This can influence stellar composition (see \citealt{2018ApJ...864..169J} and references therein) and spin \citep{2018ApJ...864...65Q}. However, even on a wider orbits at some point a planet can start to interact with the star \citep{2018A&A...618A..18R}. After a star leaves the main sequence, planets can be swallowed and digested by the red giant (see \citealt{Retter_expanding_giant, Villaver07_survive_evolution, Rodrigues15, Privitera16_interactionsII} and a review in \citealt{Veras16_postMS}). 

Statistics of planet-star interaction and its outcomes are interesting even by themselves as they are related to shaping observed distributions of planets in semi-major axes, eccentricities, and some other parameters. Furthermore, later stages of infall of a planet onto a main sequence star can be accompanied by significant energy release. Such transients  can be observed with large telescopes even at Mpc distances. \citet{metzger12} calculated energy release for different kinds of planet-star interactions and estimated the rate of such events.
These authors estimated the rate of mergers of massive planets with main sequence stars as 0.1-1 per year per galaxy. In this paper we aim to advance these estimates using a more detailed model based on population synthesis approach.

The paper is structured as follows. In the next section we describe the model we use. In Sec.~\ref{sec:results} we present our results on the rates of planet-star coalescence. Discussion is given in Sec.~\ref{sec:discus}, and we summarize our results in Sec.~\ref{sec:concl}. 

\section{Model} \label{sec:model}

In this section we describe the model we use and our assumptions about properties of stars, planets, and interaction between them. We start with the description of the evolution of individual star-planet systems. Then different types of final stages of planet-star interactions are discussed. After that we present our assumptions about planetary properties (masses, semi-major axes, densities, etc.). Finally, we describe the code we used to perform the population synthesis.

\subsection{Star-Planet System Evolution}	\label{sec:model:tidal}

In our simulation, we consider two processes that may lead to planet-star coalescence: tidal evolution and  expansion of the host star.

In a general case, tidal deformation of both star and planet leads to energy dissipation and changes in orbital parameters. The efficiency of dissipation is usually characterized by the tidal quality factor, which is proportional to the maximum energy $E_0$ stored in the tidal deformation divided by the energy lost in one cycle \citep{Goldreich_Soter}:
\begin{equation} 
	Q = 2\pi \frac{E_0}{\oint \left( -\frac{dE}{dt}\right)dt}.
\end{equation}
It is useful to introduce the modified tidal quality factor
\begin{equation}
	Q'=\frac{3Q}{2k},
\end{equation}
where $k$ is the second-order potential Love number of a star. $Q'$ is equal to $Q$ for an isotropic fluid body, for which $k=3/2$.

The evolution of the semi-major axis $a$ of a planet due to tidal dissipation within the host star is described by the following formula \citep{2004ApJ...614..955M}:
\begin{equation}
    \frac{1}{a}\frac{da}{dt} = - \frac{9}{2}\left(\frac{n}{Q'_{\star}}\right)\left(\frac{M_\mathrm{p}}{M_{\star}}\right)
    \left(\frac{R_{\star}}{a}\right)^5 \left[1 - \left(\frac{P_\mathrm{orb}}{P_\mathrm{rot}}\right)\right],
    \label{eq:adot_beautiful}
\end{equation}
where $n$ is the mean orbital motion, $M_\star$, $R_{\star}$, $Q'_{\star}$ are the mass, radius, and modified tidal quality factor of the star, $M_\mathrm{p}$ is the mass of the planet, $P_\mathrm{orb}$ and $P_\mathrm{rot}$ are the periods of orbital motion and stellar rotation, correspondingly.
This equation can be rewritten as
\begin{equation}
   \frac{da}{dt} = -\frac{9}{2} \sqrt{\frac{G}{M_{\star}}} \frac{M_\mathrm{p}R_{\star}^5}{Q'_{\star}} a^{-11/2}
    \left[1 - \left(\frac{a}{a_\mathrm{sync}}\right)^{3/2}\right],
    \label{eq:adot_work}
\end{equation}
where $a_\mathrm{sync}$ is the radius of the star-synchronous orbit. Throughout the paper, we assume that orbits of all planets are circular\footnote{See the discussion on this assumption in \autoref{sec:discus:other}.} and their spin period is equal to the orbital one. For these reasons, we neglect the tidal dissipation within the planet. Thus, in order to characterize the tidal evolution of a star-planet system with a given masses in this framework, one needs to know the values of $Q'_{\star}$ and the evolution of $R_{\star}$ and $P_\mathrm{rot}$ (or, in other terms, $a_\mathrm{sync}$).

Since there is a wide diversity of mechanisms leading to the tidal dissipation, the tidal quality factor depends in a complex way on the key parameters of the system: orbital period, masses and radii of the star and the planet, etc. However, some statistical properties of exoplanets can be well reproduced using the same value of the tidal quality factor for all systems. For example, \citet{Jackson08} explained variations of the eccentricity distribution with semi-major axis by tidal damping of eccentricity of close-in planets. The best agreement of their model with observations is achieved with the stellar quality factor $Q'_{\star}=10^{5.5}$. At the same time, many other authors (e.~g.~\citealt{Penev12_Q}) constrain $Q'_{\star}$ to the values $>10^{7}$. In order to investigate the relation between the merger rate and the tidal quality factor, we perform our calculations for a set of $\log Q'_{\star}$~=~5.5, 6.0, 6.5, 7.0, 7.5, and 8.0. The value of $Q'_{\star}$ is assumed to be constant during the main sequence (MS) phase. When a star leaves the MS, its $Q'_{\star}$ is switched to $10^{5.5}$ independently of its mass. A value close to this was obtained by \cite{2017ApJ...849L..11W} for the subgiant WASP-12. Note also, that during the post-MS evolution the main reason of planet engulfment is stellar expansion, not tides, so the value of $Q'_{\star}$ on this stage is less important (see, however, \citealt{2009ApJ...705L..81V, 2013ApJ...772..143S}).

The second ingredient of the system evolution is evolution of the stellar radius. For a star with a given mass and given age the radius is calculated in our code by interpolation of the grid of PARSEC (PAdova and TRieste Stellar Evolution Code) evolutionary tracks \citep{Bressan12_tracks,Chen14_tracks} for the solar metallicity $Z=0.02$.

Thirdly, the rotation period of a star determines the sign of the orbital radius derivative and also has an influence on its value. The rotational evolution of stars is still a debatable issue (see e.~g.~\citealt{2018ApJ...862...90G}). To date there is no commonly accepted theory that explains the change of stellar rotation periods. So we use a relatively simple approach based on the following considerations. Low-mass stars with convective envelopes undergo significant slowdown during the main sequence phase due to angular momentum loss via magnetized stellar wind. More massive stars have radiative envelopes, thus their winds are relatively inefficient in terms of angular momentum losses (except very massive stars), and so their rotation periods do not change significantly (see e.~g.~Table~1 in \citealt{2000A&A...361..101M}). Thus, we use two different approaches for these two classes of stars.

If the stellar mass is less than $1.25 M_{\sun}$, then on the zero-age main sequence (ZAMS) we set its spin period to $2.5P_\mathrm{rot,cr}$, where
\begin{equation}
    P_\mathrm{rot,cr} = 2\pi\sqrt{\frac{3R_{\star}^3}{2GM_{\star}}}.
\end{equation}
I.~e., the rotation velocity at stellar equator is 0.4 of the critical velocity, as in \cite{2012A&A...537A.146E} (of course, this is a necessary simplification; initial spin periods can be influenced by several agents, including planets, see \citealt{2018A&A...619A..80G}). During the MS stage rotational evolution is described by the following equation \citep{2010ApJ...722..222B}:
\begin{equation}
    \frac{dP_\mathrm{rot}}{dt} = \left(\frac{k_\mathrm{I}P_\mathrm{rot}}{\tau_\mathrm{c}} + \frac{\tau_\mathrm{c}}{k_\mathrm{C}P_\mathrm{rot}}\right)^{-1},
\end{equation}
where $k_\mathrm{I}=452$~Myr~day$^{-1}$ and $k_\mathrm{C}=0.646$~days~Myr$^{-1}$ are numerically calculated coefficients, and $\tau_\mathrm{c}$ is the convective turnover timescale. We interpolate the tabulated values of $\tau_\mathrm{c}$ from \cite{2010ApJ...721..675B}. These values are in a good agreement with more recent calculations of \cite{2011AN....332..897M}. The turnover timescale remains practically constant during the MS  phase, as shown by \citealt{1996ApJ...457..340K} (see their Fig.~3).

For stars more massive than $1.25 M_{\sun}$ we calculate the spin period in a different way. We use a correlation between stellar angular momentum and its mass, based on the results by \citet{Kawaler87_rot} and \citet{Paz-Chinchon15_rot_kepler}:
\begin{equation}
	J_\star = 
    \begin{cases} 
		2 \times 10^{49} \, \mathrm{g} \, \mathrm{cm}^2 \, \mathrm{s}^{-1} \left( \frac{M_\star}{M_{\sun}} \right)^{4.9}, & 1.25M_{\sun} < M_{\star} < 1.7M_{\sun} \\
		9 \times 10^{49} \, \mathrm{g} \, \mathrm{cm}^2 \, \mathrm{s}^{-1} \left( \frac{M_\star}{M_{\sun}}\right)^{2.1}, & M_\star\geq 1.7 M_{\sun}
	\end{cases}
\end{equation}
The spin period is then calculated using the relations:
\begin{equation}
	P_\mathrm{rot} = \frac{2\pi}{\Omega_\mathrm{rot}} = 2\pi\frac{I_{\star}}{J_{\star}} \approx 2\pi\frac{2M_{\star}R_{\star}^2}{5J_{\star}},
\end{equation}
where $\Omega_\mathrm{rot}$ is the angular speed of stellar rotation, $I_{\star}$ is the stellar moment of inertia, and we assume that the star is a homogeneous, uniformly rotating sphere. For these stars ($M_{\star}>1.25 M_{\sun}$) the spin period is calculated at the ZAMS and does not change with time. 

\begin{figure}
	\centering
	\includegraphics[width=\linewidth]{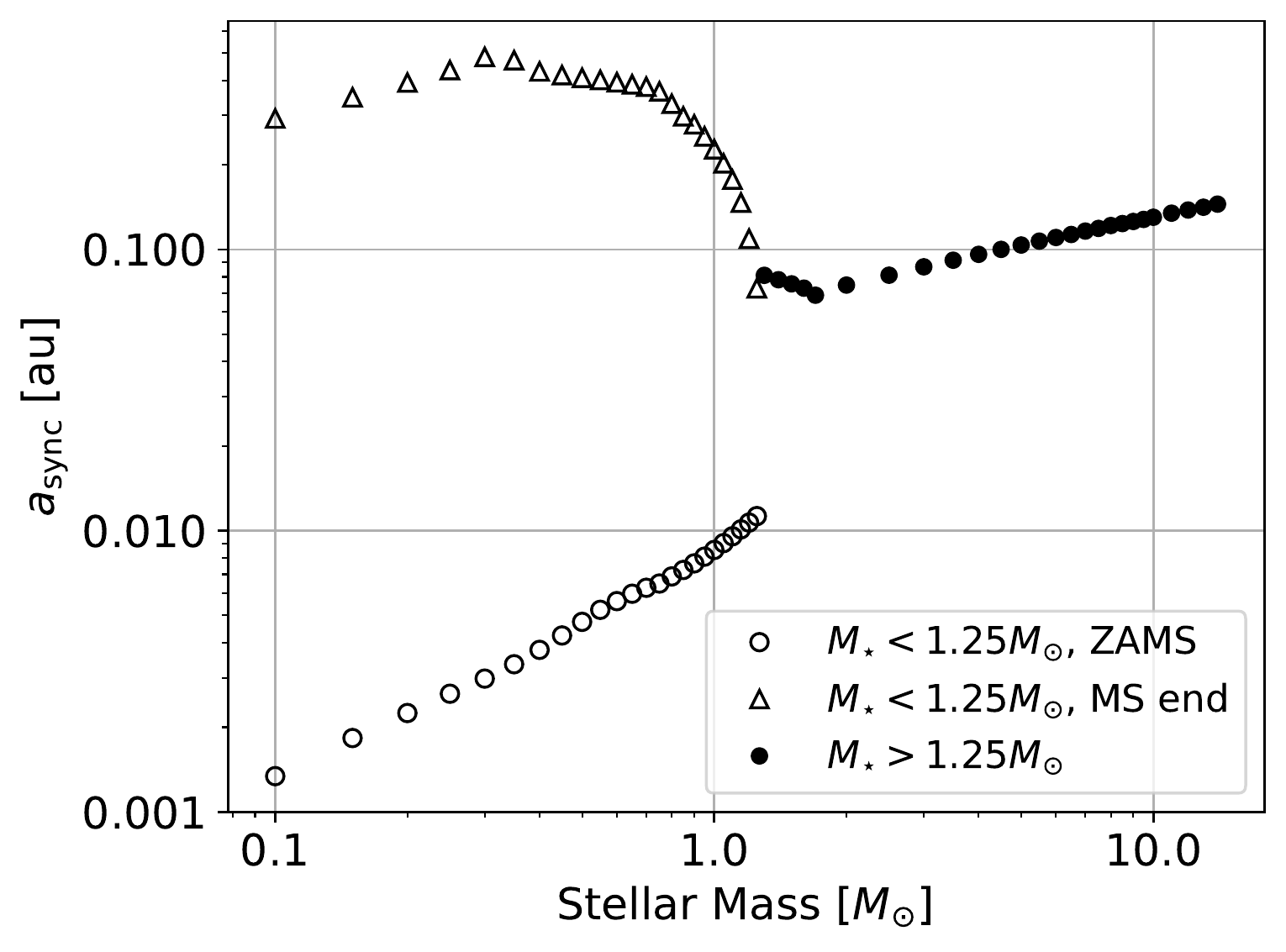}
	\caption{Radius of the star-synchronous orbit ($a_\mathrm{sync}$) vs. stellar mass for the mass range used in our modeling. For $M_{\star}<1.25M_{\sun}$, the spin rate of the star evolves along the main sequence; the values for the zero-age main sequence (ZAMS) and the end of the MS phase are shown. For more massive stars the spin rate is assumed to be constant (see text for details).}
	\label{fig:a_sync}
\end{figure}

The plot of $a_\mathrm{sync}$ for the whole range of stellar masses used in the simulation is shown in Figure~\ref{fig:a_sync}.  

\subsection{Types of mergers and their luminosities}	\label{sec:model:lum}

\citet{metzger12} classified possible planet-star interactions into three main categories. Here we follow this approach. Correspondingly, we accept the luminosity estimates provided by these authors.

The outcome of interaction mainly depends on the ratio of average densities of the star, $\bar \rho_\star$, and the planet, $\bar \rho_\mathrm{p}$. 

If the planet has lower density than the star then mass transfer is stable, and so it proceeds on a long time scale. This situation might not result in a significant luminosity enhancement. Below we use the term {\it stable accretion} for this case.

In the range of relative densities $1\lesssim \bar \rho_\mathrm{p}/\bar \rho_\star \lesssim 5$ tidal forces destroy the planet. This is {\it tidal disruption}. Thus, on a dynamical timescale the whole planet is transformed into an accretion disk. As a result, the accretion rate is super-Eddington and a significant optical transient with luminosity $L\sim10^{37}$~--~$10^{38}$~erg~s$^{-1}$ can appear for a time period from weeks to months.  

Finally, when the planet is very dense with respect to the star -- $\bar \rho_\mathrm{p}/\bar \rho_\star \gtrsim 5$, -- it merges with the star without tidal disruption or Roche lobe overflow. In \citet{metzger12} notation this is called {\it direct impact}. Following them, we assume that a direct impact happens if the following condition is satisfied:
\begin{equation}
	a_\mathrm{t} < R_\star + X_\mathrm{t},
    \label{eq:di_condition}
\end{equation}
where $a_\mathrm{t}\approx 2R_\star (\bar\rho_{\star}/\bar\rho_\mathrm{p})^{1/3}$ is the Roche limit, $X_\mathrm{t} \approx 0.7 (M_\mathrm{p}/M_\star)^{1/3} a_\mathrm{t}$ is the distance between the center of the planet and the Lagrangian point $L_1$, and $\bar\rho_{\star}$, $\bar\rho_\mathrm{p}$ are the mean densities of the star and the planet, correspondingly.

In this case, according to \citet{metzger12}, at first a transient visible in extreme UV or/and soft X-rays is formed with a peak luminosity $\lesssim 10^{36}$~ergs~s$^{-1}$. Then, weeks or months later an optical transient with $L\sim10^{37}$~--~$10^{38}$~ergs~s$^{-1}$ might be visible. Duration of the optical flare is about few days. 

We assume the following equations for the peak luminosity of optical transients:
\begin{equation}
	L_\mathrm{peak,di} \approx \left(4\pi\sigma\right)^{1/3} T_\mathrm{rec}^{4/3} \left(\frac{G M_{\star}}{R_{\star}}\right)^{1/3} \left(\frac{0.1 M_\mathrm{p}}{m_\mathrm{p}}\mathrm{Ry}\right)^{2/3}  
    \label{eq:l_di}
\end{equation}
for a direct impact and
\begin{equation}
	L_\mathrm{peak,td} \approx 3 \times 10^{34}\, \text{erg s}^{-1}\frac{M_\mathrm{p}}{M_{\earth}}
    \label{eq:l_td}
\end{equation}
for a tidal disruption. Here $M_{\star}$, $R_{\star}$ are the mass and radius of the star, $M_\mathrm{p}$ -- mass of the planet, $\sigma$ -- Stefan-Boltzmann constant, $T_\mathrm{rec}=6000$~K -- the recombination temperature of hydrogen, $m_\mathrm{p}$ -- proton mass, and Ry=13.6~eV. In both equations it is assumed that the star is on the main sequence. Equation~(\ref{eq:l_di}) was derived by \citet{metzger12}, while equation~(\ref{eq:l_td}) is a fit to their numerical results.

At the post-MS evolutionary stages, when most of planet-star mergers occur, practically all of them belong to the direct impact class due to low average stellar density. Their luminosity must be smaller then in the case of the host star on the main sequence, because both the density of the outer layers of the star and the velocity of the perturbing planet are smaller. We estimate the upper limit of the luminosity of such events as the minimum of two quantities: $L_\mathrm{peak,di}$ given above and
\begin{equation}
    L_\mathrm{giant} = \frac{3}{4}C_\mathrm{d}\eta G^{3/2}R_\mathrm{p}^2 M_{\star}(M_{\star}+M_\mathrm{p})^{3/2}R_{\star}^{-9/2},
    \label{eq:l_giant}
\end{equation}
taken from  \cite{2018ApJ...853L...1M}. Here $R_\mathrm{p}$ is the planetary radius, $C_\mathrm{d}=1$ is a dimensionless drag coefficient, $\eta=0.1$ is the ratio of the stellar envelope density to the average density of the entire star.

\subsection{Initial distribution in the $a_0-M_\mathrm{p}$ plane}	\label{sec:model:ma}

\begin{figure}
	\centering
	\includegraphics[width=\linewidth]{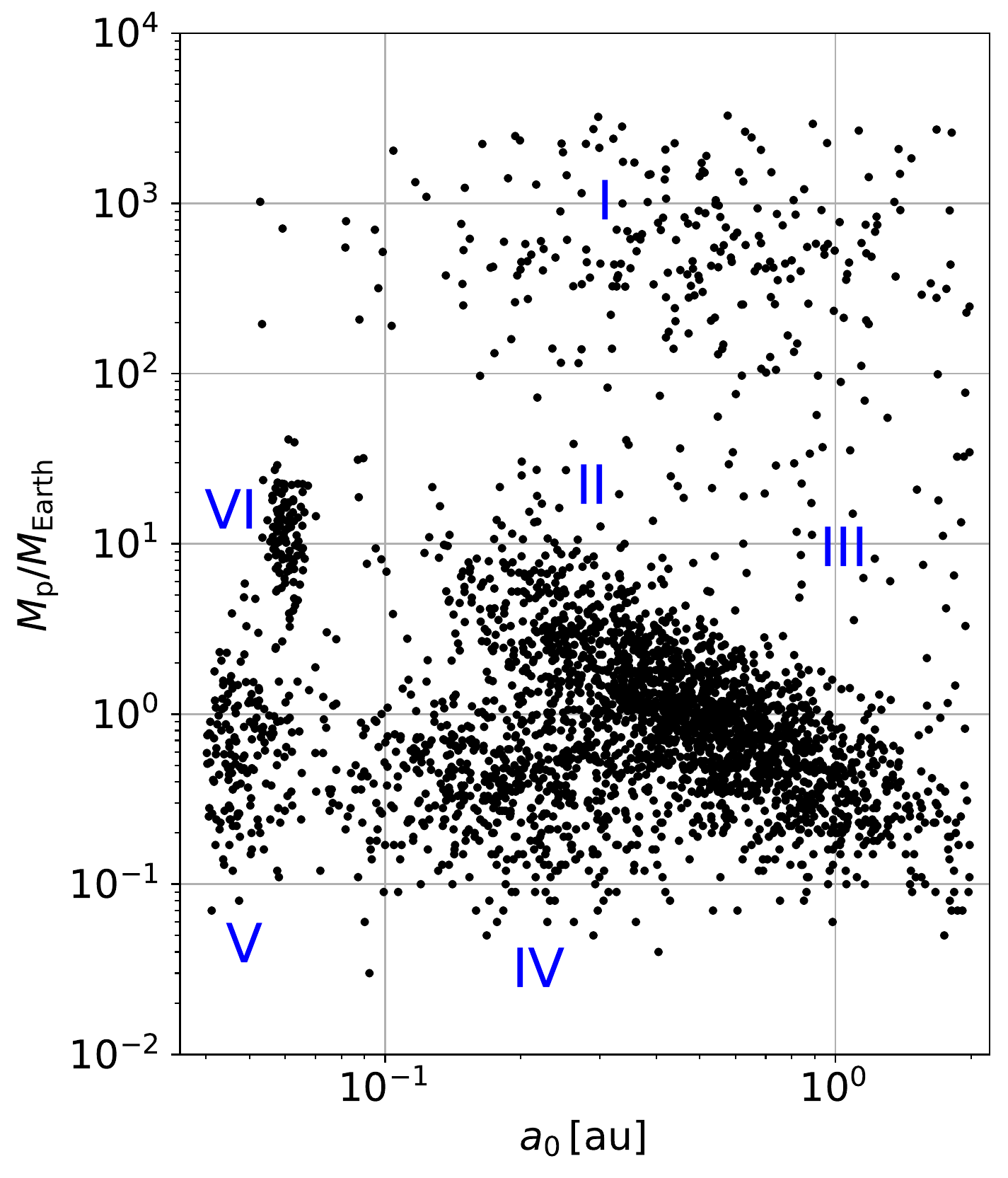}
	\caption{Initial mass -- semi-major axis distribution of planets.
    Data are obtained from our analytical fit of detailed calculations by \citet{Alibert13} (see text for details).
    Number of planets shown in the diagram is 3000. Groups of planets are enumerated according to Table~\ref{tab:distribution}. Note, that the group III is not numerous, but covers a wide range of parameters.}
	\label{fig:initial_ma}
\end{figure}

Initial distribution of planets in semi-major axis, $a_0$, and mass, $M_\mathrm{p}$, is one of the most important ingredients of our model. Selection effects do not allow to derive a precise $a_0-M_\mathrm{p}$ distribution from observations for the whole range of parameters. Thus, in our simulation we choose a different approach. We start with results of population synthesis by \citet{Alibert13} (see their Fig.~5). We select several representative groups of planets in the $a_0-M_\mathrm{p}$ plane and fit each of these groups by an analytical distribution. We use the following distributions:
\begin{itemize}
	\item Two-dimensional uniform distribution (in logarithmic scale).
    \item Two-dimensional log-normal distribution. This is the product of two one-dimensional distributions with probability density function
    	\begin{equation}
			p(x)= \frac{1}{x\sqrt{2\pi\sigma^2}} \exp \left( -\frac{(\ln x -\zeta)^2}{2\sigma^2} \right),
		\end{equation}
        where $\zeta$ and $\sigma$ are the parameters of the distribution.
    \item Bivariate Gaussian distribution in log scale:
		\begin{equation}
			\begin{split}
				p(x,y) = \frac{1}{xy\, 2\pi \sigma_\mathrm{x} \sigma_\mathrm{y} \sqrt{1-\rho^2}} \times \\ 
    			\exp \left( -\frac{\phi^2/\sigma_\mathrm{x}^2 + \psi^2/\sigma_\mathrm{y}^2 - 2\rho \phi\psi/\sigma_\mathrm{x}\sigma_\mathrm{y}}{2(1-\rho^2)} \right),
    		\end{split}
		\end{equation}
		where we denote $\phi = \log x - \zeta_\mathrm{x}$, $\psi = \log y - \zeta_\mathrm{y}$ for brevity (note that here we use the decimal logarithm instead of the natural logarithm in the log-normal distribution). Five parameters define this distribution: $\zeta_\mathrm{x}$, $\zeta_\mathrm{y}$, $\sigma_\mathrm{x}$, $\sigma_\mathrm{y}$, and $\rho$; the latter is the correlation coefficient.
\end{itemize}
Parameters for each group are given in Table~\ref{tab:distribution}, and the resulting distribution is shown in Figure~\ref{fig:initial_ma}.

\begin{table}
	\centering
	\caption{Parameters of $a_0-M_\mathrm{p}$ distribution. See text for symbols definitions.}
	\label{tab:distribution}
	\begin{tabular}{cclc} 
		\hline
        \hline
		Group & Distribution & Parameters & \% of total \\
		\hline
        \hline
		\multirow{4}{*}{I} & \multirow{4}{*}{2D Log-normal} & $\zeta_\mathrm{a}=\ln 0.5$ & \multirow{4}{*}{7.9} \\
        & & $\zeta_\mathrm{M}=\ln 500$ & \\
        & & $\sigma_\mathrm{a}=0.9$ & \\
        & & $\sigma_\mathrm{M}=1.0$ & \\
		\hline
        \multirow{5}{*}{II} & \multirow{5}{*}{Bivariate Gaussian in log} & $\zeta_\mathrm{a}=\log 0.5$ & \multirow{5}{*}{55.0} \\
        & & $\zeta_\mathrm{M}=0$ & \\
        & & $\sigma_\mathrm{a}=0.25$ & \\
        & & $\sigma_\mathrm{M}=0.45$ & \\
        & & $\rho = -0.8$ & \\
        \hline
		\multirow{4}{*}{III} & \multirow{4}{*}{Uniform in log} & $\log a_\mathrm{min} = -0.7$ & \multirow{4}{*}{6.1} \\
        & & $\log a_\mathrm{max} = 0.4$ & \\
        & & $\log M_\mathrm{min} = -1$ & \\
        & & $\log M_\mathrm{max} = 1.6$ & \\
		\hline
        \multirow{4}{*}{IV} & \multirow{4}{*}{2D Log-normal} & $\zeta_\mathrm{a}=\ln 0.2$ & \multirow{4}{*}{20.7} \\
        & & $\zeta_\mathrm{M}=\ln 0.4$ & \\
        & & $\sigma_\mathrm{a}=0.5$ & \\
        & & $\sigma_\mathrm{M}=0.8$ & \\
		\hline
        \multirow{4}{*}{V} & \multirow{4}{*}{2D Log-normal} & $\zeta_\mathrm{a}=\ln 0.045$ & \multirow{4}{*}{7.1} \\
        & & $\zeta_\mathrm{M}=\ln 0.7$ & \\
        & & $\sigma_\mathrm{a}=0.2$ & \\
        & & $\sigma_\mathrm{M}=0.8$ & \\
		\hline
        \multirow{4}{*}{VI} & \multirow{4}{*}{2D Log-normal} & $\zeta_\mathrm{a}=\ln 0.06$ & \multirow{4}{*}{3.2} \\
        & & $\zeta_\mathrm{M}=\ln 12$ & \\
        & & $\sigma_\mathrm{a}=0.05$ & \\
        & & $\sigma_\mathrm{M}=0.5$ & \\
		\hline
        \hline
    \end{tabular}
\end{table}

Tidal evolution is not important for distant planets ($a\gtrsim 0.1$ au). Also, planets with large semi-major axis cannot be consumed by red giants ($a \gtrsim$~few au). So, we do not fit planets from \citet{Alibert13} if  $a > 2$ au (see discussion in \autoref{sec:discus:giants}). The inner boundary of semi-major axis distribution is set to $0.04$ au, similar to the typical inner boundary of a protoplanetary disk (see discussion and reference in \citealt{2016ApJ...823...58M}; of course, for real disks this value can vary). For the maximum planetary mass we adopt a conventional value of 13 Jupiter masses. 

The most important class of planets from the point of view of bright flares due to mergers is represented by so-called {\it hot jupiters}, i.e. gas giants with short orbital periods. According to \citet{Wright12_occurrence}, the occurrence rate of hot jupiters with mass larger than $0.1M_\mathrm{Jup}$ and period less than 10 days is $1.2 \pm 0.4$ \%. This estimates in confirmed also my more recent studies \citep{2018haex.bookE.155B, 2018haex.bookE.154S}. Our distribution provides the occurrence rate of such planets within this range.

We assume that the distribution of planetary mass and initial orbital radius has the same form for stars of any mass, varying only the number of planets in a system (eq.~\ref{eq:n_pl}). We discuss this assumption below in~\autoref{sec:discus:initial}.

\subsection{Properties of Planets and Systems}

\begin{figure}
	\centering
	\includegraphics[width=\linewidth]{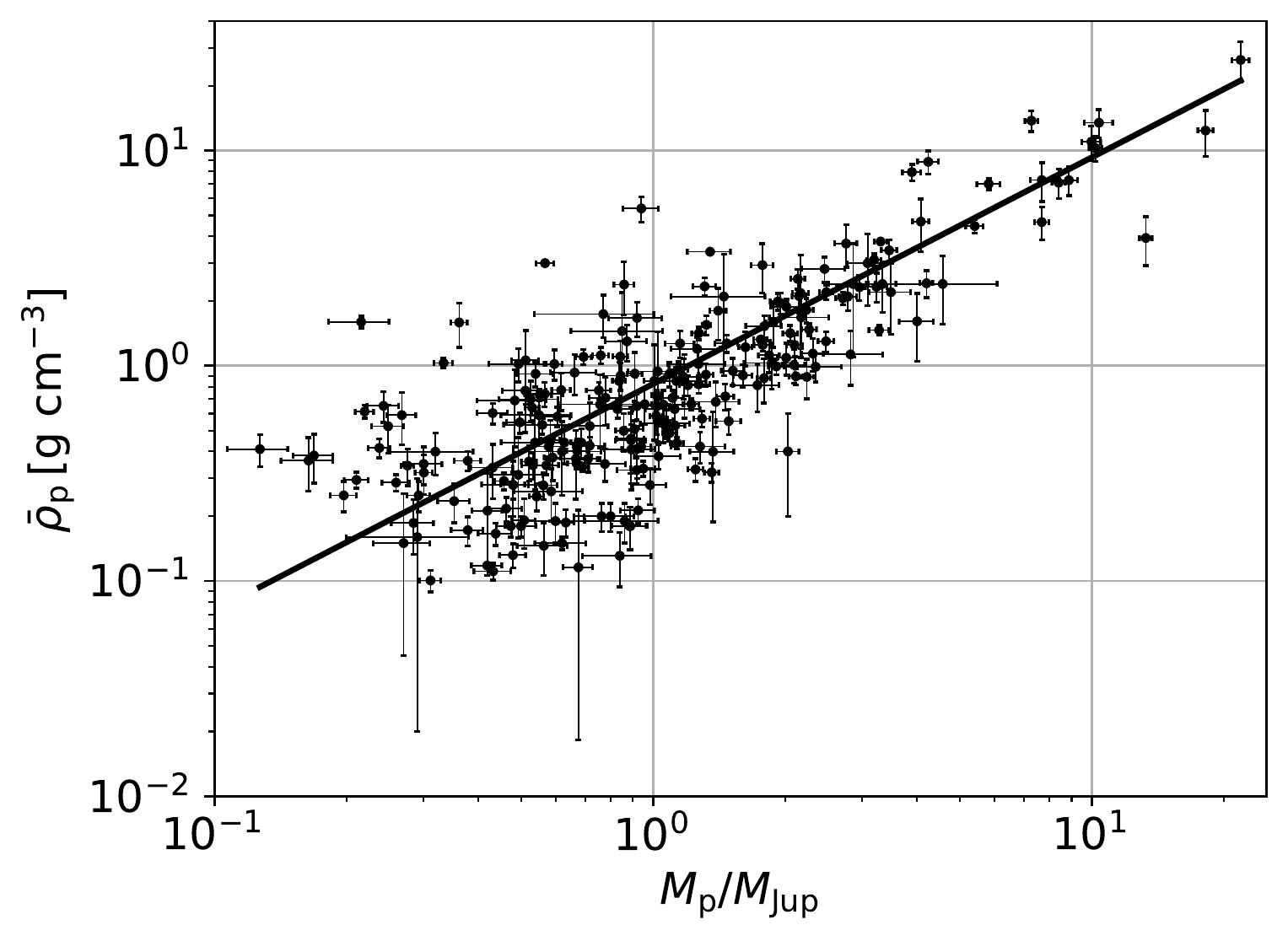}
	\caption{Planetary density vs. planetary mass. Data are taken from \url{http://exoplanets.org/}, thick solid line corresponds to the best fit, see eq. (\ref{eq:rhop}).}
	\label{fig:density}
\end{figure}

To calculate the output of a planet-star interaction, in addition to the mass and initial orbital radius of the planet one should know its density. We calculate the planetary density according to:
\begin{equation} 
	\bar\rho_\mathrm{p} = 
    \begin{cases}
		\bar\rho_{\earth}, & M_\mathrm{p}\leq M_{\earth} \\
		\bar\rho_{\earth} \left( \frac{M_\mathrm{p}}{M_{\earth}} \right)^{-0.46}, & M_{\earth}<M_\mathrm{p} \leq 200M_{\earth} \\
		1.9 \times 10^{-3} \left(\frac{M_\mathrm{p}}{M_{\earth}}\right)^{1.05}\mathrm{g}\,\mathrm{cm}^{-3}, & M_\mathrm{p} > 200M_{\earth}
	\end{cases}
    \label{eq:rhop}
\end{equation}
In the first two cases ($M_\mathrm{p} \leq 200M_{\earth}$) functional dependences are taken from \citet{Laughlin15_exogeophys}. For larger masses we fit the data from the Exoplanets Data Explorer database\footnote{\url{http://exoplanets.org/}, see \cite{2011PASP..123..412W}.} (see~\autoref{fig:density}).

Some properties of planetary systems in our model are set by the mass of the host star. They include the number of planets in the system $n_\mathrm{pl}$ and the protoplanetary disk lifetime $t_\mathrm{disk}$ (see subsection~\ref{sec:code} for the role these parameters play in the simulation). For the first one, we construct the following relations, based on discussion by \citet{Alibert11}:
\begin{equation}
	n_\mathrm{pl}(M_{\star}) = 
    \begin{cases}
		\left(\frac{M_{\star}}{M_{\sun}}\right)^{1.2} n_\mathrm{pl}(M_{\sun}), & M_{\star} < 1.5 M_{\sun} \\
		6.5, & M_{\star} \geq 1.5 M_{\sun},
	\end{cases}
	\label{eq:n_pl}
\end{equation}
where $n_\mathrm{pl}(M_{\sun}) = 4$ is the number of planets in the range of semi-major axes $0.04\,\text{au} \leq a \leq 2\,\text{au}$ in the Solar system (namely, Mercury, Venus, Earth and Mars).

For $t_\mathrm{disk}$ we use a value of 10~Myr as a conservative upper estimate (e.~g. \citet{2007ApJ...662.1067H} found that the fraction of stars with protoplanetary disk falls to practically zero at 10~Myr age). For stars with $M_{\star} \geq 1.5\, M_\odot$, we adopt the inverse square root dependence on stellar mass following \citet{Alibert11}:
\begin{equation}
	t_\mathrm{disk} = 
    \begin{cases}
		10^7 \; \mathrm{yr}, & M_{\star}<1.5M_{\sun} \\
		10^7 \left(\frac{M_{\star}}{1.5M_{\sun}}\right)^{-1/2} \mathrm{yr}, & M_{\star} \geq 1.5M_{\sun}.
	\end{cases}
    \label{eq:tdisk}
\end{equation}

\subsection{Stellar Initial Mass Function and Star Formation History}

To calculate the rate of coalescences in a galaxy\footnote{Below we consider only conditions for the Milky Way, and so the rate refers to our Galaxy.} basing on behavior of individual planetary systems, one needs information about properties of galactic stellar population. In our model we neglect variations of metallicity, i.e. we use a single value $Z=0.02$. Thus,  we need  to specify just the number of stars (depending on their ages) and their mass distribution.

We adopt stellar initial mass function (IMF) from \citet{Kroupa01}:
\begin{equation}
	P(M<M_{\star}<M+dM) \propto 
    \begin{cases} 
		M^{-1.3}, & M < 0.5 M_{\sun} \\
		M^{-2.3}, & M \geq 0.5 M_{\sun}
	\end{cases}
    \label{eq:kroupa_imf}
\end{equation}
and set the range of stellar mass from 0.09 to 14 $M_{\sun}$. More massive stars live too short to allow planet formation and evolution. 

We assume that the IMF does not change, but the star formation rate (SFR) depends on time:
\begin{equation} 
	\mathrm{SFR}(\tau) = 
    \begin{cases}
		3\, M_{\sun}/\mathrm{yr}, & \tau \leq 7 \times 10^9\, \mathrm{yr} \\
		0, & 7 \times 10^9\, \mathrm{yr} < \tau \leq 9.5 \times 10^9\, \mathrm{yr}\\
		10\, M_{\sun}/\mathrm{yr}, & 9.5 \times 10^9\, \mathrm{yr} < \tau \leq 12.5 \times 10^9\, \mathrm{yr}
	\end{cases}
    \label{eq:sfr_steps}
\end{equation}
Here $\tau$ is lookback Galactic time ($\tau=0$ corresponds to the present epoch).
This is a simplified form of star formation history presented in \citet{Haywood16_sfr}.

\subsection{Description of the Code}
\label{sec:code}

The model described above is used as a part of population synthesis code. Here we briefly describe it.

We assume that planet-planet interactions do not play significant role in the orbital evolution. This allows us to consider evolution of each planet independently. By Monte-Carlo method, $N=10^7$ star-planet pairs are generated. In our model, such pairs are characterized by three initial parameters: $M_\star$, $M_\mathrm{p}$, $a_0$. Masses and initial semi-major axes of planets are generated using the distribution described in~\autoref{sec:model:ma}. Stellar masses are distributed according to the IMF from eq.~(\ref{eq:kroupa_imf}) multiplied by the function $n_\mathrm{pl}(M_{\star})$, eq.~(\ref{eq:n_pl}). This reflects the fact that in our treatment every star should appear in several iterations the number of which is proportional to $n_\mathrm{pl}$ for this particular star (this mimics multi-planetary systems in which each planet can be treated separately).

To calculate the effective number of stars $N_{\star}$, corresponding to $N$ iterations, we set the weight $1/n_\mathrm{pl}$ for each iteration:
\begin{equation} 
	N_{\star} = \sum_{i=1}^{N} \frac{1}{n_\mathrm{pl, i}}.
    \label{eq:nstar}
\end{equation}
For each generated star-planet pair its evolution is followed. The tidal migration starts at the system age equal to the maximum of two times: the beginning of the main sequence stage of the host star and the protoplanetary disk dissipation time $t_\mathrm{disk}$ (we do not model the pre-main sequence evolution and use the parameters on the zero-age main sequence as initial conditions). After calculation of the system evolution, the program determines whether a coalescence occurs in the pair. If yes -- then its time, type, evolutionary stage of the star, and (if the coalescence produces a transient) the peak luminosity are added to the collected statistics of events. A merger is regarded as occurred if the orbital radius of the planet becomes smaller than the limiting value $a_\mathrm{lim}$. If the condition~(\ref{eq:di_condition}) is satisfied then $a_\mathrm{lim} = R_{\star}$ and a direct impact happens; else, we have $a_\mathrm{lim} = a_\mathrm{t}$ and so a tidal disruption or stable accretion takes place. The latter two types are distinguished by the ratio of planetary and stellar mean density as described in~\autoref{sec:model:lum}.

Based on calculations for $N$ systems, the probability distribution of mergers in age $p(t)$ is calculated. It is equal to the number of mergers in a sample during a small time interval, divided by the duration of the interval and by the effective number of stars, $N_{\star}$, in the sample. Then total coalescence rate in the Galaxy is computed by summing contributions from stellar populations of different ages according to the galactic star formation history SFR (in units stars per year) weighted by $p(t)$:
\begin{equation} 
	n(t) = \int\limits_0^t \mathrm{SFR}(\tau)\,p(\tau)\,d\tau.
    \label{eq:sfr_conv}
\end{equation}
One may calculate $p(t)$ for a selected type of mergers as well as for all of them, and so to derive rates of different phenomena.

\section{Results}	\label{sec:results}

\subsection{Evolution of a Single Planet}

\begin{figure}
	\centering
	\includegraphics[width=\linewidth]{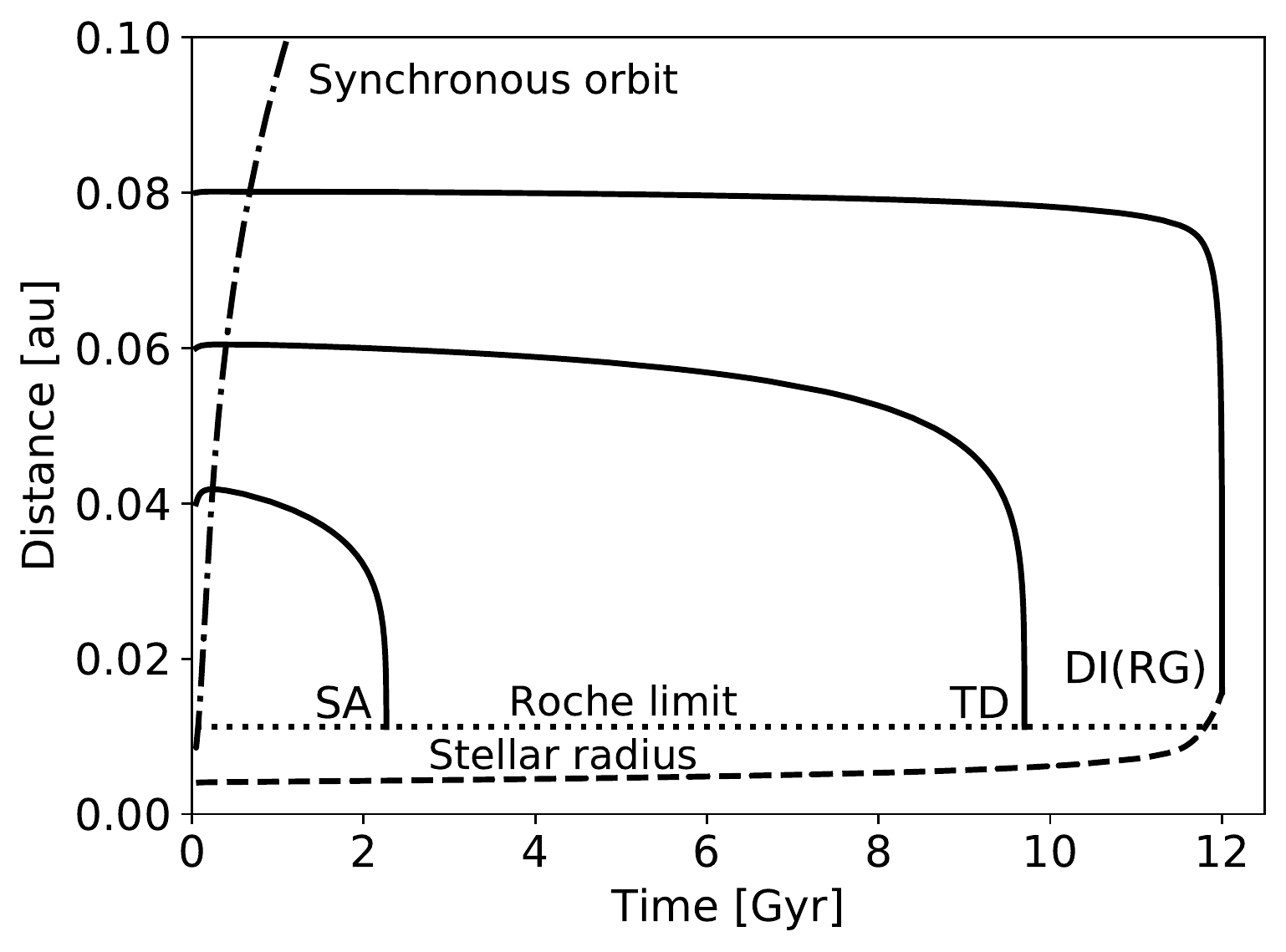}
	\caption{Tidal evolution of planets with different initial semi-major axes (solid lines) for $\log Q'_{\star}=5.5$ during the whole lifetime of the host star. Near each merger point its type is noted: SA -- stable accretion, TD -- tidal disruption, DI(RG) -- direct impact at the red giant stage. Stellar radius is plotted by dashed line; it is stopped at the moment of last planet's fall. Dotted and dashed-dot line mark the values of Roche limit ($a_\mathrm{t}$) and radius of the star-synchronous orbit ($a_\mathrm{sync}$), correspondingly. Here $M_\star = 1 M_{\sun}$, $M_\mathrm{p} = 1 M_\mathrm{Jup}$.}
	\label{fig:single_change_a0}
\end{figure}

\begin{figure}
	\centering
	\includegraphics[width=\linewidth]{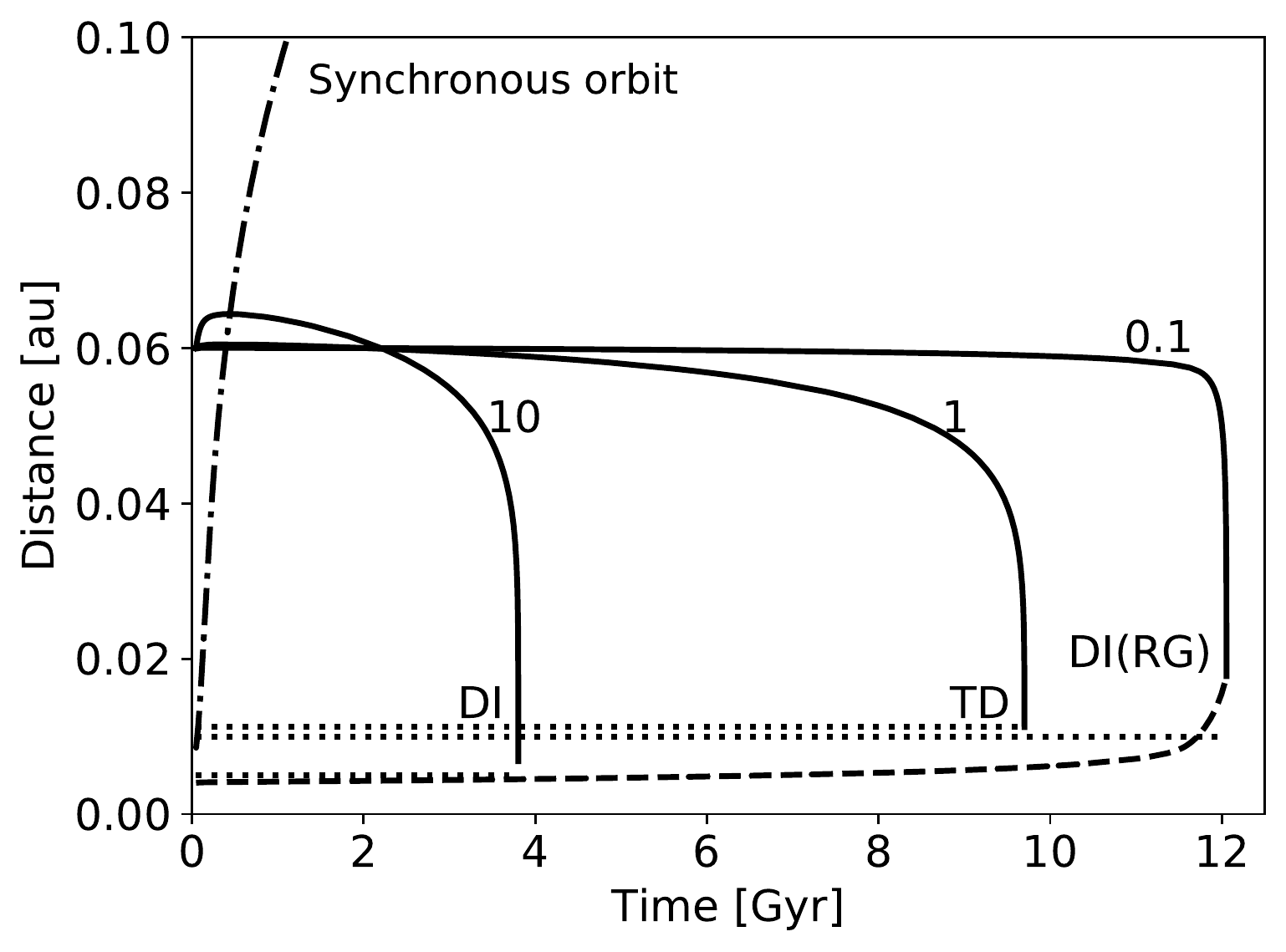}
	\caption{Tidal evolution of planets of different masses (solid lines) around a solar mass star for $\log Q'_{\star}=5.5$ during the whole lifetime of the host star. Near each solid curve the planetary mass in units of Jupiter mass and the type of merger are marked (DI -- direct impact, TD -- tidal disruption, DI(RG) -- direct impact at red giant stage). Stellar radius is plotted by dashed line, it is terminated at the moment of last planet's fall. Dashed-dot line marks the radius of the star-synchronous orbit ($a_\mathrm{sync}$), and dotted lines -- Roche limits ($a_\mathrm{t}$) for each planetary mass. }
	\label{fig:single_change_mp}
\end{figure}

\begin{figure}
	\centering
	\includegraphics[width=\linewidth]{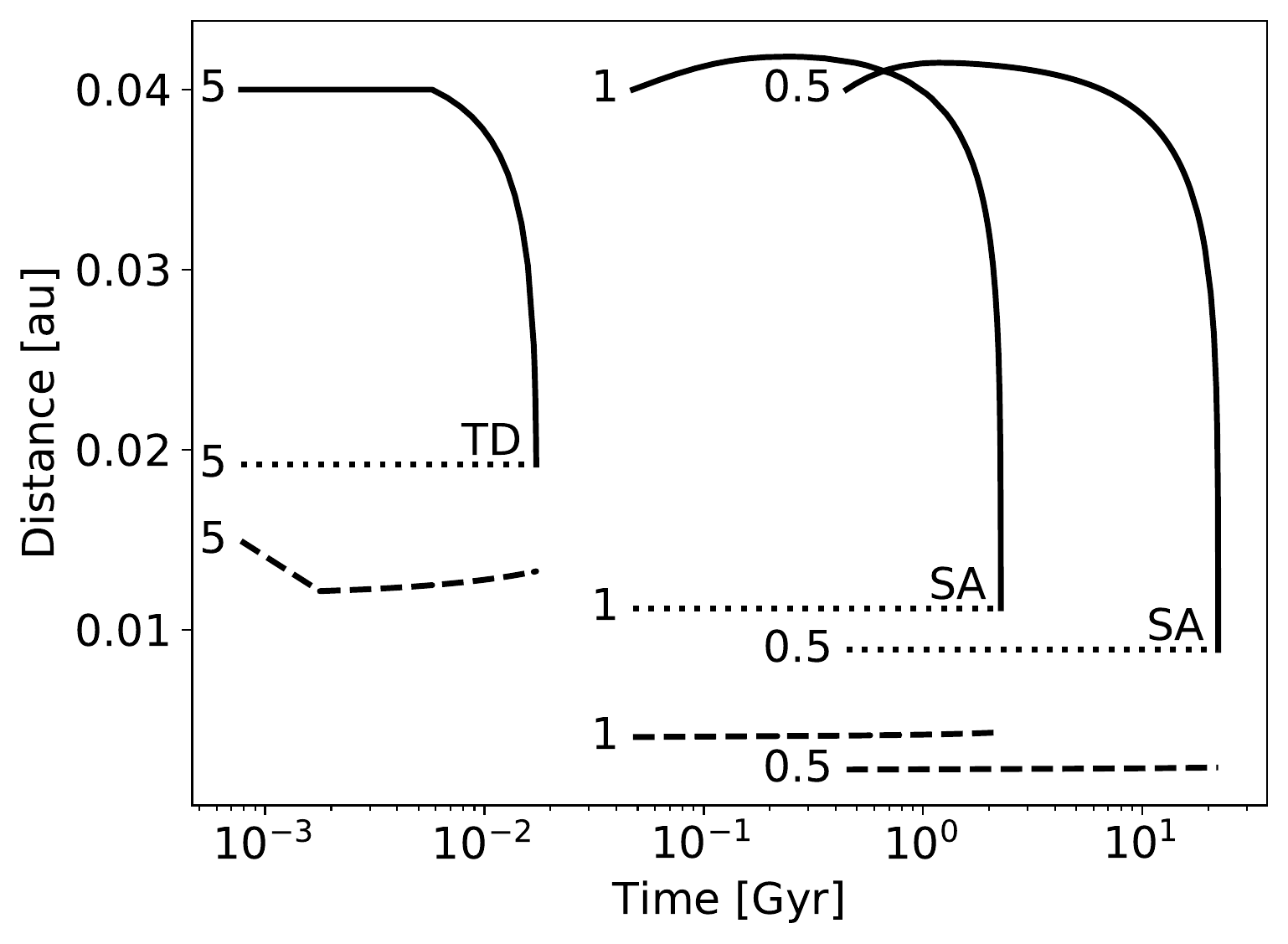}
	\caption{Tidal evolution of Jupiter-mass planets around stars of different masses (solid lines) for $\log Q'_{\star}=5.5$ during the whole lifetime of the host star. Stellar radii are plotted by dashed lines, and Roche limits ($a_\mathrm{t}$) for each star-planet pair  -- by dotted lines. Each curve is marked by a corresponding stellar mass in solar mass units. All the curves for each pair begin at the moment of the beginning of the planet migration (see subsection~\ref{sec:code}) and are terminated at the moment of merger. Near each merger point its type is indicated (SA -- stable accretion, TD -- tidal disruption). }
	\label{fig:single_change_ms}
\end{figure}

At first, we present examples of orbital evolution calculated using our model for different initial conditions (Figures~\ref{fig:single_change_a0}--\ref{fig:single_change_ms}). Of course, similar calculations have been done many times by different authors, and we present them for illustrative purposes only.

In these examples masses and initial orbital radii are chosen such that coalescences in systems occur at ages less than the age of the Galaxy (except the case of a $0.5M_{\sun}$ star and a Jupiter-mass planet in Fig~\ref{fig:single_change_ms}). Types of mergers are indicated in the corresponding plots.

Note, that for a low-mass star (red dwarf) a planet might be massive enough (several Jupiter masses) to coalesce within the Galactic lifetime even for the lowest tidal quality factor $Q'_{\star}=10^{5.5}$. This is one of the reasons why we do not expect to have many mergers of rocky and icy planets with red dwarfs.

\subsection{Probability Distribution of Mergers}
\label{sec:results:distr}

In this section, we present the statistics of outcomes of $N$ simulated star-planet pairs evolution. For $N=10^7$ the corresponding value of effective number of stars $N_{\star} \approx 6.72 \times 10^6$. That is, the average occurrence of planets in our model is approximately 1.5 planets per star. This is caused by prevalence of low-mass stars, which have less planets according to our assumptions.

For our study, the most important property of the sample of star-planet pairs is the distribution of mergers in time since the formation of the systems (i.e., a system's age at the moment of coalescence). There are three types of mergers that lead to observable transients: direct impacts during the main sequence and post-main sequence stages and tidal disruptions at the main sequence stage. 

\begin{figure}
	\centering
	\includegraphics[width=\linewidth]{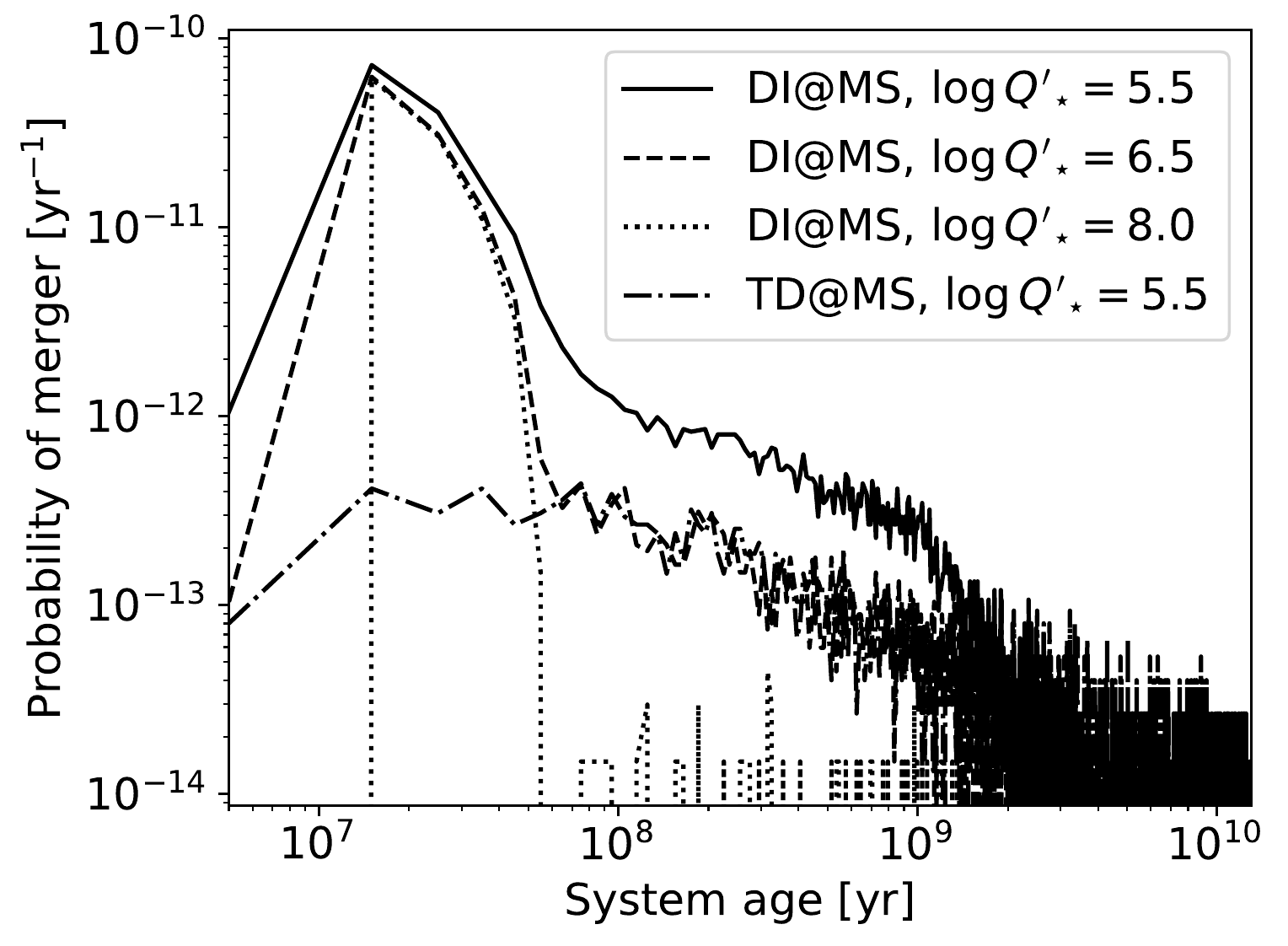}
	\caption{Probability $p(t)$ of a merger of a planet with a main sequence host star vs. system's age for different values of $Q'_{\star}$ and two merger types: direct impact (DI) and tidal disruption (TD). See text for details.}
	\label{fig:dif}
\end{figure}

For direct impacts and tidal disruptions during the main sequence stage the temporal distributions are shown in Figure~\ref{fig:dif}. For direct impacts, results for three values of $Q'_{\star}$ are presented; for TD, only the curve for the minimal $Q'_{\star}$ is plotted, because in case of higher $Q'_{\star}$ the values of $p$ are too uncertain. Results in Fig.~\ref{fig:dif} correspond to a population of stars with the IMF specified above born simultaneously in a single star formation burst. Each curve represents the probability of coalescence (direct impact or tidal disruption) for an average star in this population at a given time. In other words, this is the coalescence rate (of each type) per star. Curves at large ages are distorted by statistical noise. One can see that the distribution peaks at ages about few $\times \, 10^7$ years and decreases by orders of magnitude in $\sim 10^9$ years. It means that young systems make the most significant contribution to the rate of mergers with main sequence stars. To calculate this rate, one needs to know accurately the star formation history just during past billion years or so.

The feature of the curves for direct impacts is that the rate at ages less than about 50 million years is very weakly affected by the change of $Q'_{\star}$. This is an evidence that most of the mergers occurring in first tens million years after star formation have non-tidal nature. These mergers are ingestions of the innermost planets in the process of the host star expansion at the main sequence stage. We will demonstrate it below.

The temporal distribution for direct impacts at post-main sequence stages (not shown in Fig.~\ref{fig:dif}) has a maximum at system age $\sim 3\times 10^{7}$~years; the peak value is an order of magnitude higher than for direct impacts on main sequence. After maximum, it decreases approximately by power law with an index about~$-0.8$. That means, the rate of planet engulfment during post-main sequence stages is orders of magnitude higher than the rate of planet-star mergers during the main sequence stage, which is expected also from general considerations. The relative contributions of different types of coalescences to the observable transients statistics is discussed in the following sections.

\begin{figure}
	\centering
	\includegraphics[width=\linewidth]{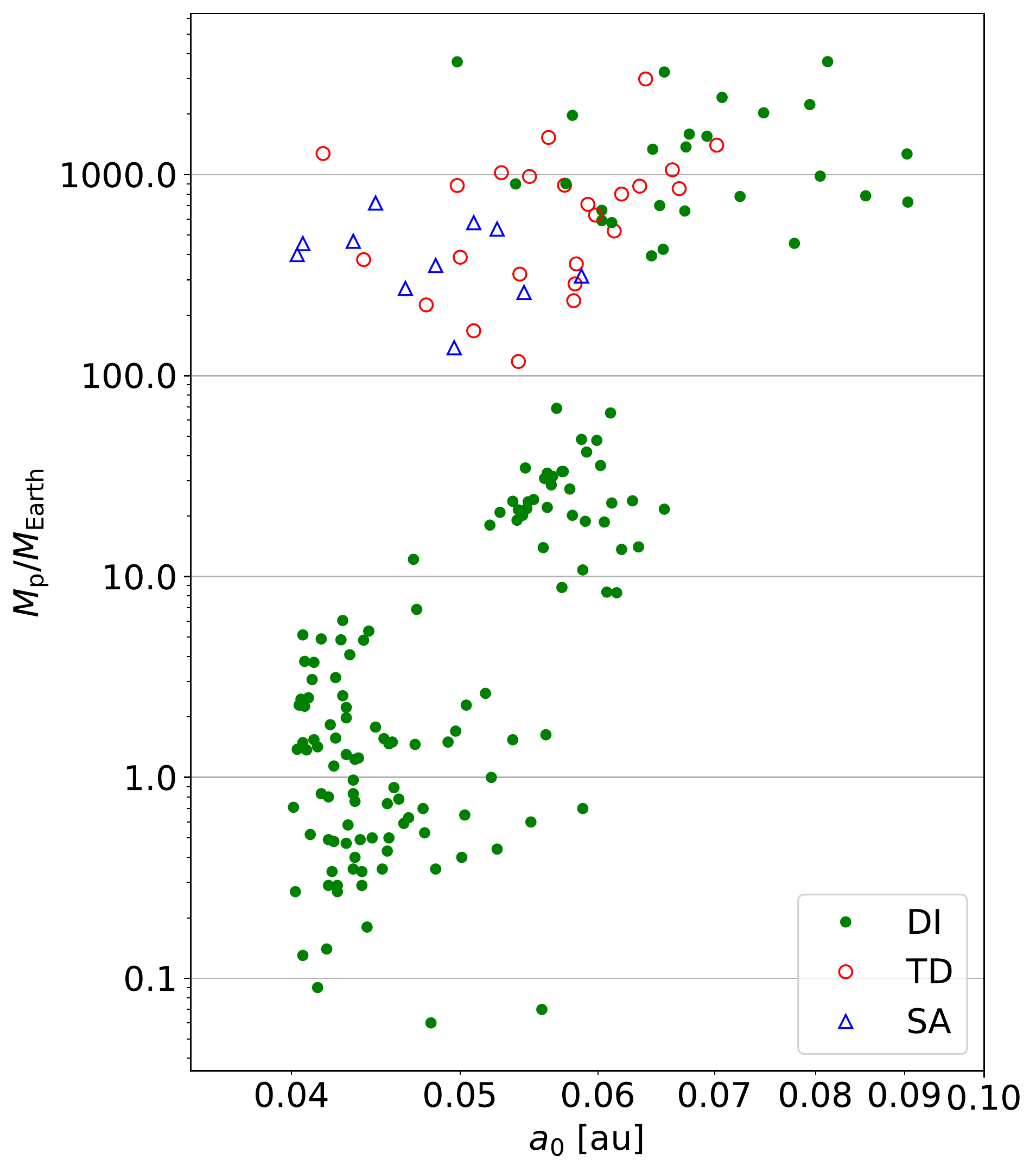}
	\caption{Mass and initial orbital radius of planets which merge with their host stars during the main sequence phase in simultaneously formed population of planetary systems in case of $\log Q'_{\star}=5.5$. Symbol style marks the type of mergers: dots -- direct impacts~(DI), circles -- tidal disruptions~(TD), triangles -- stable accretion~(SA) events. Only mergers happening at ages less than the age of the Galaxy are included. Total number of planets in the population is $10^5$, the number of planets on this diagram is 185.}
	\label{fig:fallenMS_types}
\end{figure}

As well as the age distribution of mergers, the relation between their probability and the initial system parameters is worth a discussion. Fig.~\ref{fig:fallenMS_types} shows a distribution similar to the one in Fig.~\ref{fig:initial_ma}, but only for planets which merge with their host stars during the main sequence phase at ages less than the Galactic age if $\log Q'_{\star}=5.5$.\footnote{Total numbers of planets in Fig.~\ref{fig:initial_ma} and Fig.~\ref{fig:fallenMS_types} are not equal. They are given in the figure captures.}  This diagram demonstrates that, firstly, only planets with $a \lesssim 0.1$ au satisfy these conditions (note that horizontal axis scales in Figures~\ref{fig:initial_ma}~and~\ref{fig:fallenMS_types} are different). Secondly, direct impacts dominate among low-mass planets, and practically all mergers of other types occur with hot jupiters. This is an imprint of our model of planetary density: low-mass planets (rocky and icy) have higher density than gas giants and rarely fill their Roche lobe above the stellar surface.

\begin{figure*}
    \centering
    \includegraphics[width=0.32\linewidth]{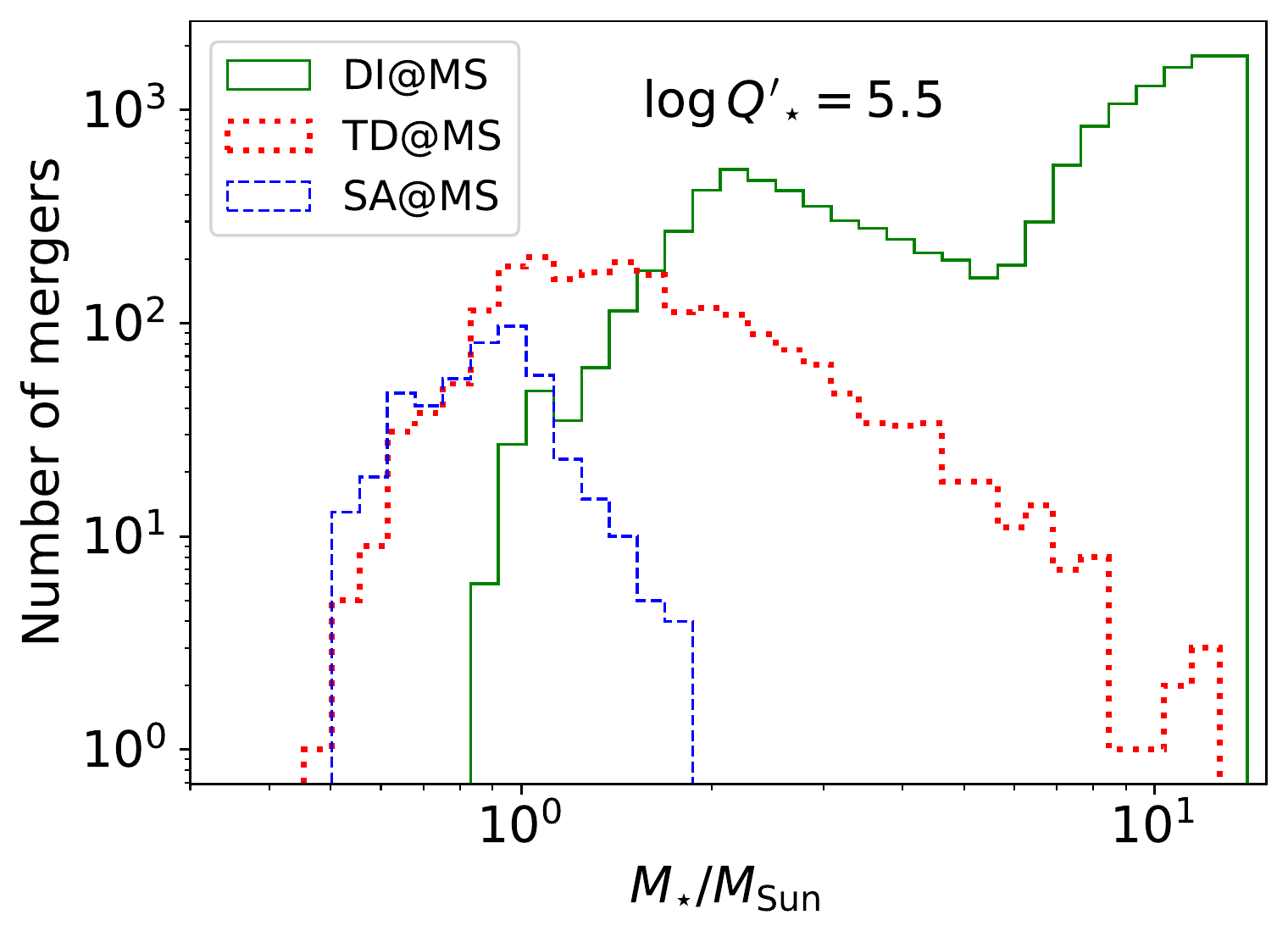}
    \includegraphics[width=0.32\linewidth]{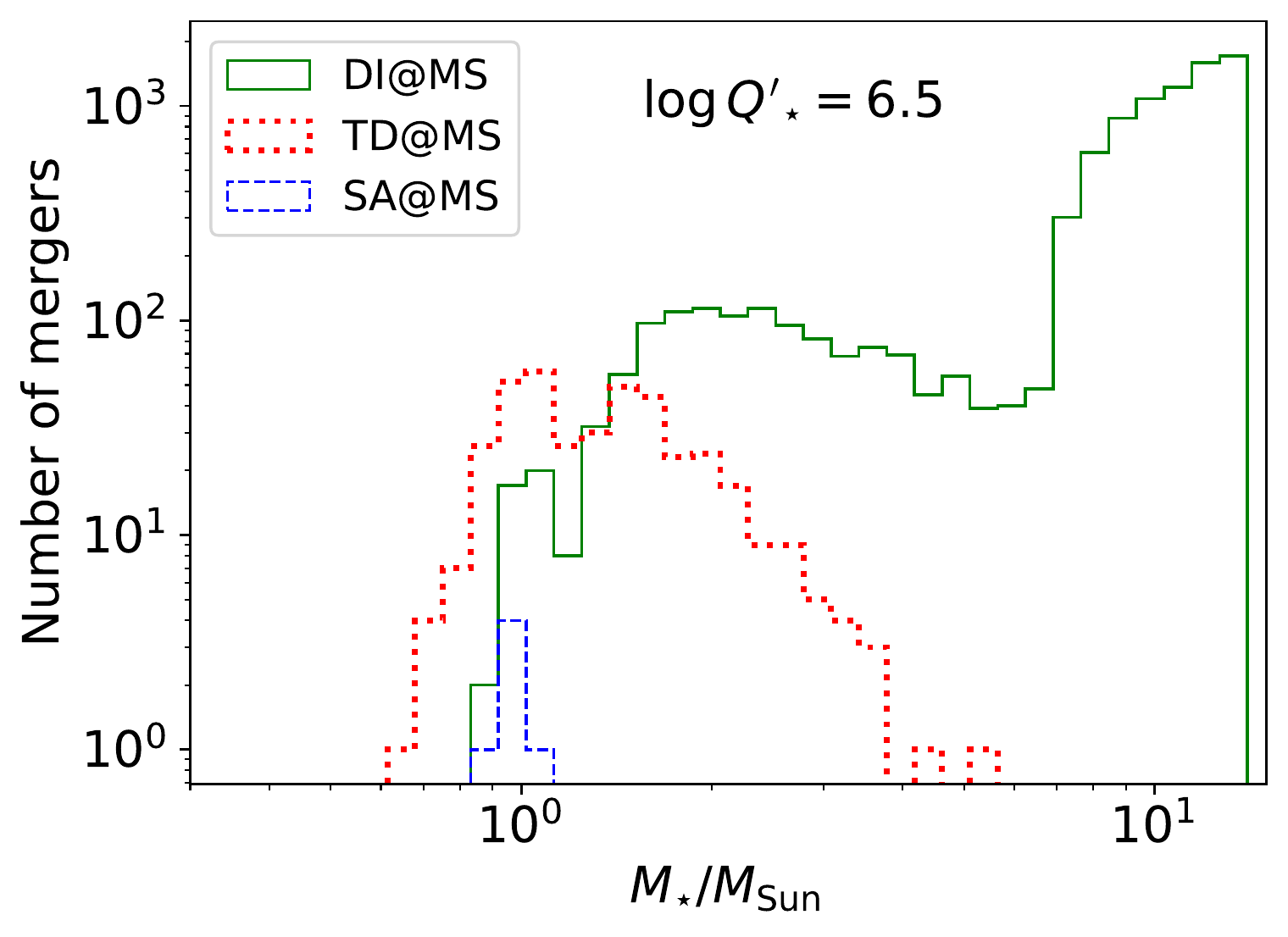}
    \includegraphics[width=0.32\linewidth]{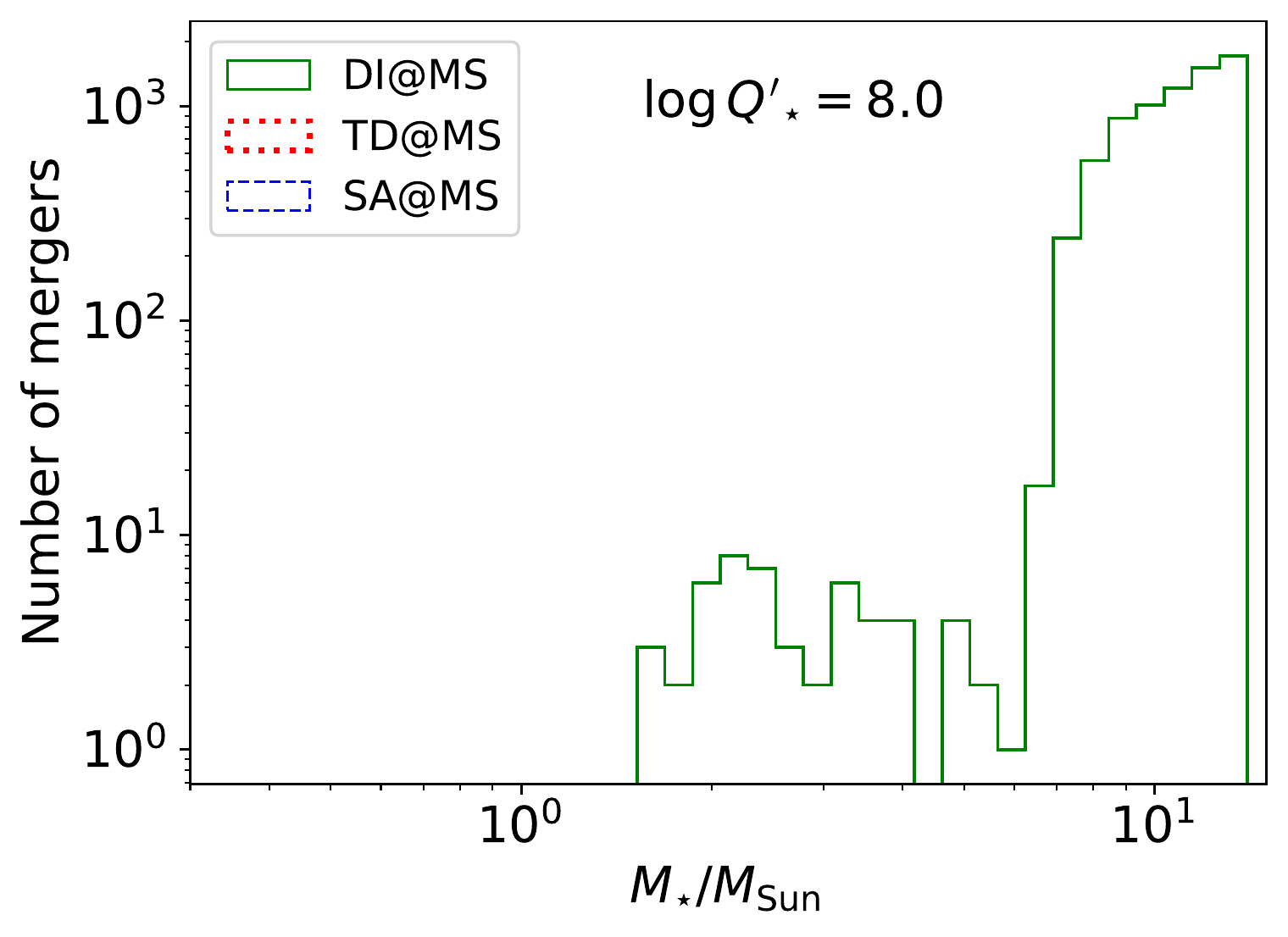}
    \caption{Distribution in mass of the host star of the mergers in the simulated population of $10^7$ star-planet systems for three values of $Q'_{\star}$. Only mergers occurring during the main sequence (MS) phase of the host star were counted. Three types of mergers: direct impact (DI), tidal disruption (TD), and stable accretion (SA), are plotted separately. Note that the logarithmic histogram bins are used.}
    \label{fig:p_mstar}
\end{figure*}

The probability distribution of mergers in the third system parameter, the mass of the host star, is shown in Figure~\ref{fig:p_mstar} (again only mergers during the main sequence phase are considered). Comparing the plots for different values of the tidal quality factor, one can see that while the number of mergers with intermediate mass stars decreases dramatically with the increase of $Q'_{\star}$, the number of mergers with stars of mass $\gtrsim 6 M_{\sun}$ practically does not change. That is, the large number of non-tidal direct impacts with main sequence host stars in first few tens million years after systems formation, which is seen in Figure~\ref{fig:dif}, is related to the most massive stars in our simulation, accounting for less than 1\% of the stellar population. These stars expand significantly during their main sequence stage, ingesting close-in planets. We discuss this in details in \autoref{sec:discus:initial}.

\subsection{Mergers in Milky Way Equivalent Galaxy}  \label{sec:results:mw}
\begin{table*}
	\centering
	\caption{Rates of mergers of different types in Milky Way-like galaxy for different values of stellar quality factor. The unit is $\mathrm{yr}^{-1}$. We denote coalescences which cause optical transients with luminosity $>10^{37}$~erg~s$^{-1}$ as ``bright transients''. Upper limits are given, when no events of such type occur in the simulation.}
	\label{tab:results}
	\begin{tabular}{cccccc}
    	\hline
    	\hline
    	$\mathrm{log} Q'_{\star}$ & Stellar evolutionary stage & Direct impact & Tidal disruption & Stable accretion & Bright transients\\
    	\hline
    	\multirow{3}{*}{5.5} & Main sequence & $1.2\times 10^{-2}$ & $2.0\times 10^{-3}$ & $5.1\times 10^{-4}$ & $0.3\times 10^{-2}$ \\
        & Post-main sequence & $2.93$ & $0.3\times 10^{-3}$ & $\lesssim 10^{-6}$ & $0.9\times 10^{-2}$\\
        & Total & $2.95$ & $2.2\times 10^{-3}$ & $5.1\times 10^{-4}$ & $1.3\times 10^{-2}$ \\
        
        \hline
        \multirow{3}{*}{6.0} & Main sequence & $8.3\times 10^{-3}$ & $1.0\times 10^{-3}$ & $1.6\times 10^{-4}$ & $0.2\times 10^{-2}$\\
        & Post-main sequence & $2.94$ & $0.4\times 10^{-3}$ &  $2.9\times 10^{-6}$ & $1.1\times 10^{-2}$ \\
        & Total & $2.95$ & $1.4\times 10^{-3}$ & $1.6\times 10^{-4}$ & $1.3\times 10^{-2}$ \\
        
        \hline
    	\multirow{3}{*}{6.5} & Main sequence & $7.5\times 10^{-3}$ & $0.4\times 10^{-3}$ & $1.2\times 10^{-5}$ & $0.2\times 10^{-2}$ \\
        & Post-main sequence & $2.94$ & $0.5\times 10^{-3}$ & $\lesssim 10^{-6}$ & $1.2\times 10^{-2}$ \\
        & Total & $2.95$ & $0.9\times 10^{-3}$ & $1.2\times 10^{-5}$ & $1.4\times 10^{-2}$ \\
        
        \hline
        \multirow{3}{*}{7.0} & Main sequence & $6.9\times 10^{-3}$ & $0.05\times 10^{-3}$ & $\lesssim 10^{-6}$ & $0.08\times 10^{-2}$\\
        & Post-main sequence & $2.94$ & $0.61\times 10^{-3}$ & $\lesssim 10^{-6}$ & $1.3\times 10^{-2}$ \\
        & Total & $2.95$ & $0.66\times 10^{-3}$ & $\lesssim 10^{-6}$ & $1.4\times 10^{-2}$ \\
        
        \hline
        \multirow{3}{*}{7.5} & Main sequence & $6.5\times 10^{-3}$ & $\lesssim 10^{-6}$ & $\lesssim 10^{-6}$ & $0.03\times 10^{-2}$ \\
        & Post-main sequence & $2.95$ & $0.6\times 10^{-3}$ & $\lesssim 10^{-6}$ & $1.3\times 10^{-2}$ \\
        & Total & $2.95$ & $0.6\times 10^{-3}$ & $\lesssim 10^{-6}$ & $1.4\times 10^{-2}$\\
        
        \hline
        \multirow{3}{*}{8.0} & Main sequence & $6.2\times 10^{-3}$ & $\lesssim 10^{-6}$ & $\lesssim 10^{-6}$ & $0.01\times 10^{-2}$ \\
        & Post-main sequence & $2.94$ & $0.6\times 10^{-3}$ & $\lesssim 10^{-6}$ & $1.4\times 10^{-2}$ \\
        & Total & $2.95$ & $0.6\times 10^{-3}$ & $\lesssim 10^{-6}$ & $1.4\times 10^{-2}$\\
    	\hline
    	\hline
    \end{tabular}
\end{table*}
We calculate the rate of planet-star coalescences in the Milky Way (or other galaxy with the same mass and star-formation history). The results for each of three merger types are shown in Table~\ref{tab:results}, including rates of mergers during stellar main sequence phase, after it and in total. All the rates are calculated for a set of stellar $Q'_{\star}$. 

\begin{figure*}
	\centering
	\includegraphics[width=0.49\linewidth]{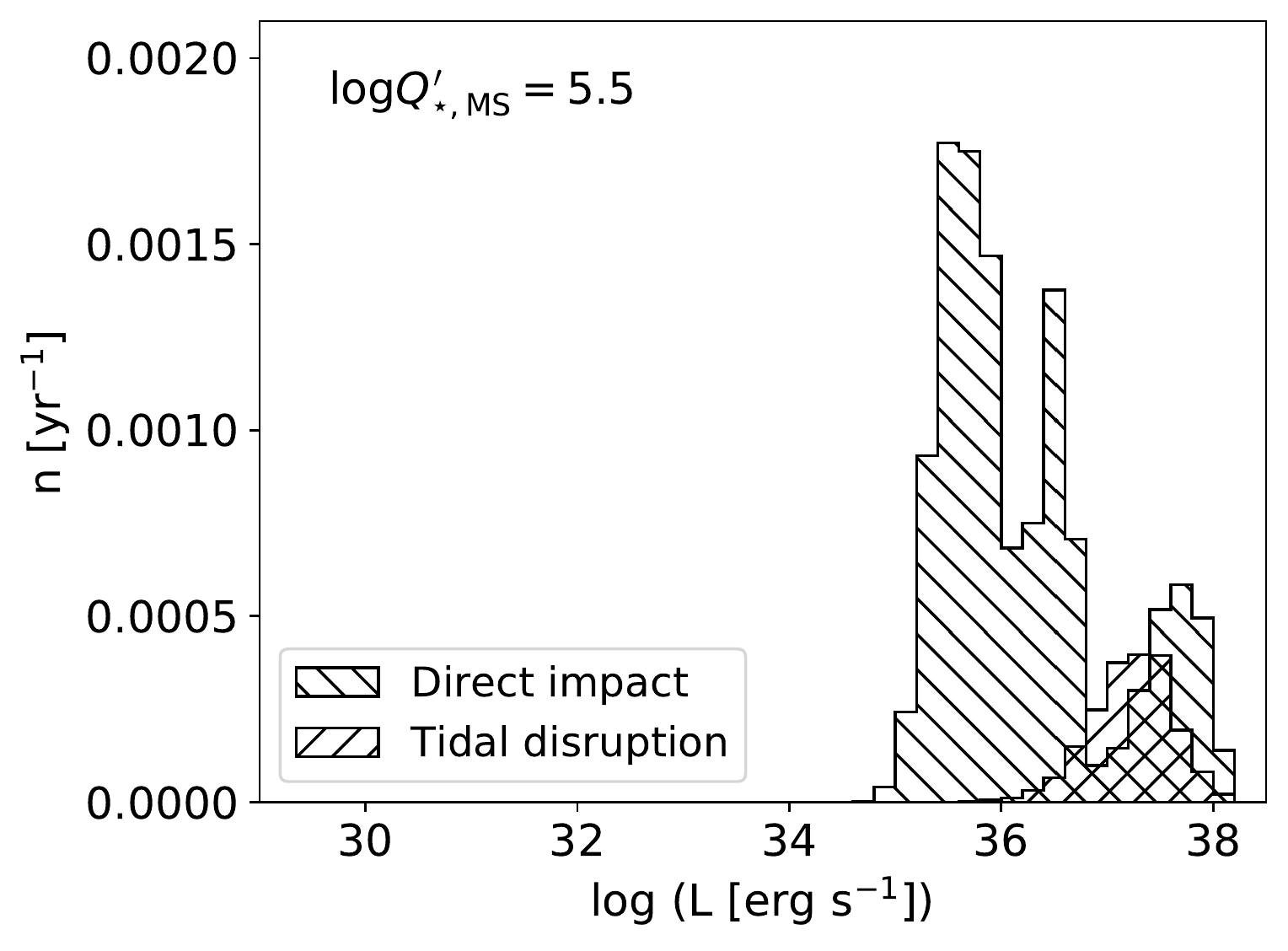}
	\includegraphics[width=0.49\linewidth]{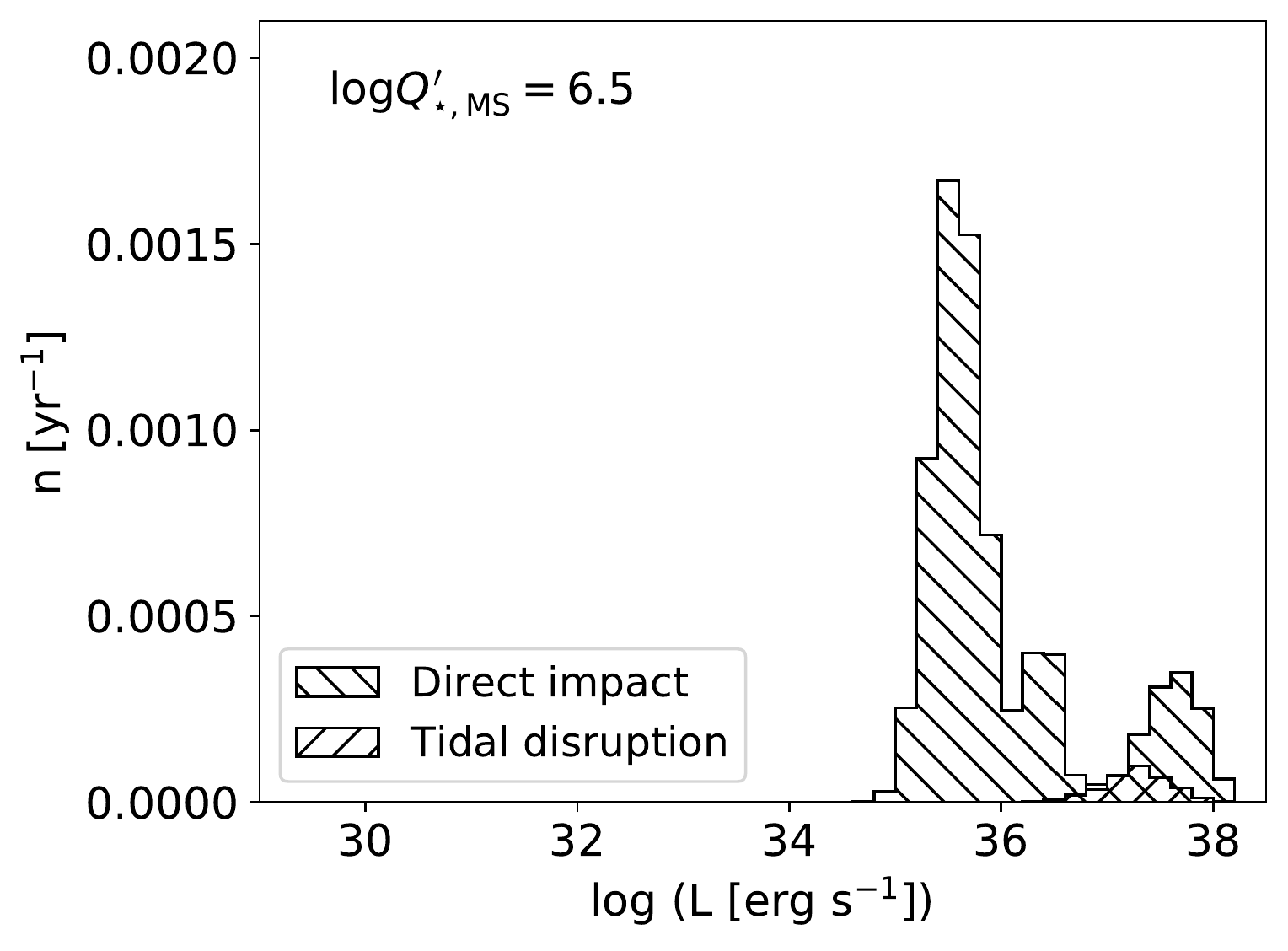}
	\includegraphics[width=0.49\linewidth]{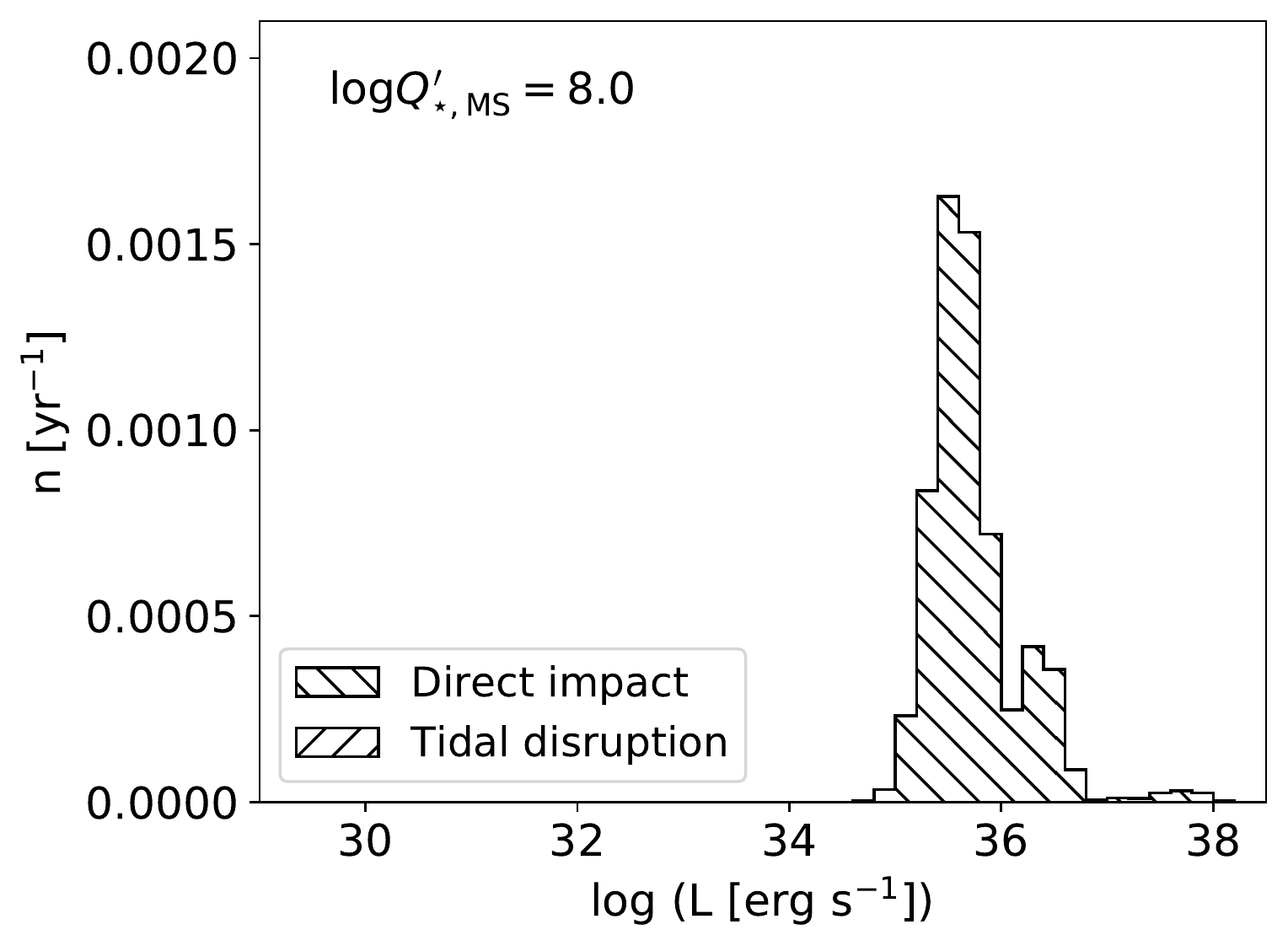}
	\includegraphics[width=0.49\linewidth]{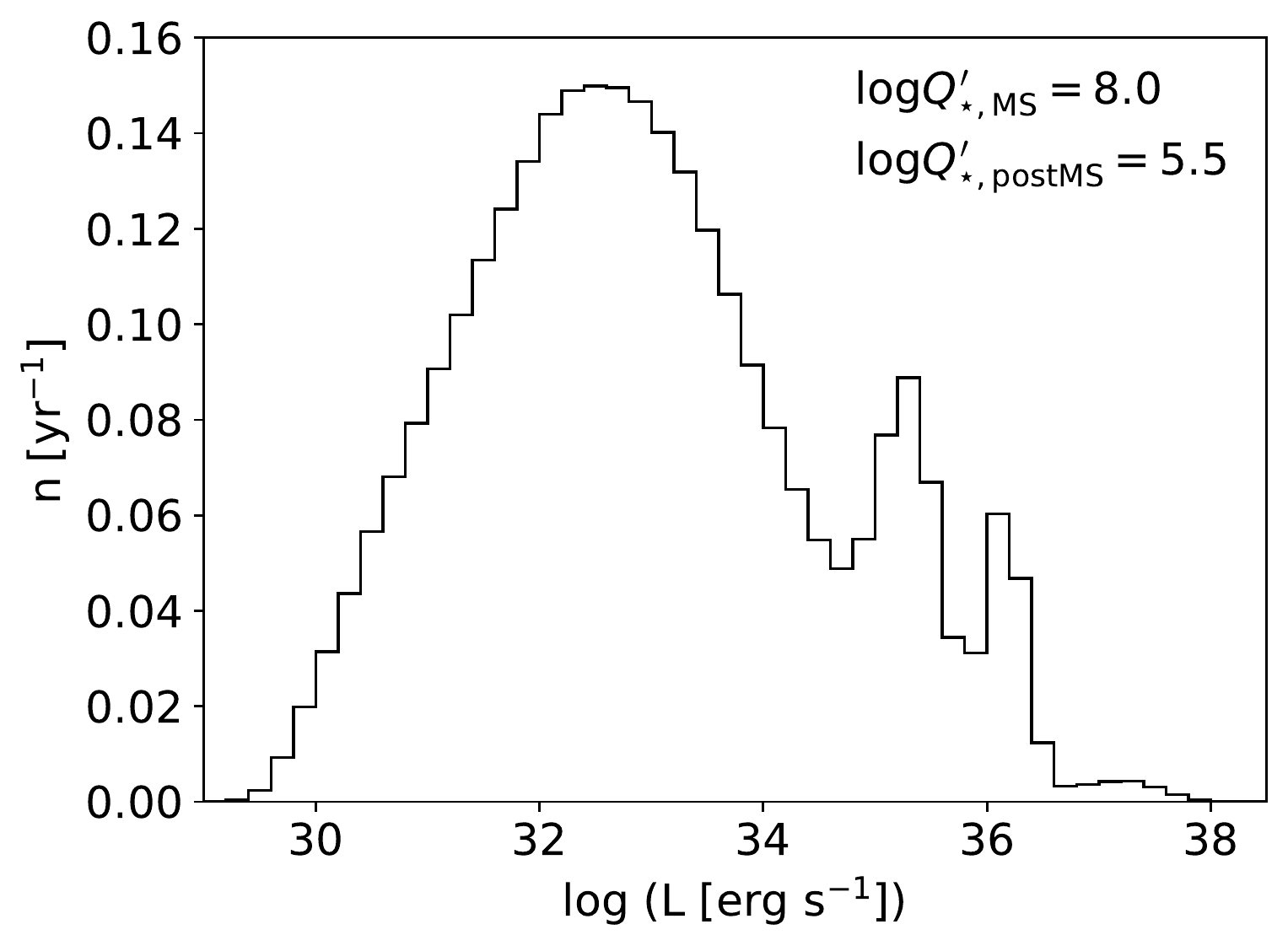}
	\caption{Distribution in luminosity of planet-star mergers in a Milky Way-like galaxy. The quantity on the ordinate is the rate of mergers per year per galaxy in a logarithmic luminosity bin. {\it Upper row and lower left plot}: mergers with main sequence host stars for different values of $Q'_{\star}$ on the main sequence (type of merger is marked by hatching). {\it Lower right plot}: mergers with post-main sequence host stars in case of $\log Q'_{\star, \mathrm{MS}}=8.0$; the curves for other values of $Q'_{\star, \mathrm{MS}}$ are practically similar. At post-MS stages $\log Q'_{\star, \mathrm{postMS}}=5.5$ for all stars in our simulation.}
	\label{fig:N_lgL}
\end{figure*}

As was already mentioned, potentially observable mergers are direct impacts and, if the host star is on the main sequence, tidal disruptions. The distribution of these transients in luminosity is plotted in Fig.~\ref{fig:N_lgL}. 

For direct impacts with main sequence stars, one can see three peaks of the luminosity distribution. This corresponds to the shape of nearby planets mass distribution, which has also three peaks (it can be seen in Fig.~\ref{fig:fallenMS_types}). Most probable events have luminosity of corresponding optical transients about $10^{35.5}$~erg~s$^{-1}$. This peak of the distribution, which looks practically the same for different $Q'_{\star}$, corresponds to already mentioned ingestions of close-in planets by expanding massive main sequence stars. The less pronounced peaks at $10^{36.5}$~erg~s$^{-1}$ and at $10^{37.5}-10^{38}$~erg~s$^{-1}$ are formed by the mergers caused mostly by tides; their rate drops off significantly with increasing of $Q'_{\star}$. The latter peak corresponds to the case considered by \citet{metzger12} -- a coalescence of a jupiter-mass planet with its host star. The Galactic rate of such events in our simulation varies from $\approx10^{-4}~\mathrm{yr}^{-1}$ for $\log Q'_{\star}=8.0$ to $\approx2\times 10^{-3}~\mathrm{yr}^{-1}$ for $\log Q'_{\star}=5.5$ (here only direct impacts are counted). As was already mentioned, \citet{metzger12} calculated that such mergers lead not only to optical transients, but also are preceded by extremely UV/soft X-ray emission with luminosity $\lesssim10^{36}$~erg~s$^{-1}$.

As for tidal disruptions, they occur with planets within smaller mass range and hence have smaller luminosity dispersion. The luminosity distribution of them peaks on~$\sim~10^{37}-10^{37.5}$~erg~s$^{-1}$. Their rate depends strongly on the stellar tidal quality factor. For main sequence host stars, it is about $2\times 10^{-3}~\mathrm{yr}^{-1}$ for $\log Q'_{\star}=5.5$, and for high $Q'_{\star}$ it is negligibly small.

In Figure~\ref{fig:N_lgL}, the luminosity distribution for mergers of planets with post-main sequence stars is also presented (lower right plot). The rate of such events is orders of magnitude higher than for the mergers during the main sequence stage, however, the distribution peaks at much lower luminosities ($10^{32}-10^{33}$~erg~s$^{-1}$). At higher luminosities, the distribution matches the distribution of the masses of close-in planets, as in the case of a host star on the main sequence.

The rates of mergers leading to transients with luminosity higher than $10^{37}$~erg~s$^{-1}$, regardless of their type, are listed in the last column of Table~\ref{tab:results}. Our results show that such transients occur in a Milky Way equivalent galaxy approximately once in 80 years. The rates of bright transients are dominated by mergers with post-MS stars. Note, however, that the luminosity of such events in our simulation is highly uncertain and may be overestimated (see discussion in \autoref{sec:discus:giants}). Taking into consideration only mergers with main sequence hosts, we obtain one event in 300 to 10,000 years, depending on the stellar tidal quality factor.

\section{Discussion}	\label{sec:discus}

In this section we discuss the possibility of observational tests of our results, compare them with previous works and dwell on the sources of uncertainties in our calculations.

\subsection{Possibility of Observations}	\label{sec:discus:obs}

Taking into account the luminosities and rates of studied events, it is likely that interstellar extinction would prevent to observe them in the Milky Way. Thus, to detect them, a deep survey of nearby galaxies is needed. \footnote{This has been already stated by \citet{metzger12}.} Such a survey is expected to be made next decade by Large Synoptic Survey Telescope (LSST).

The maximum luminosity of planet-star mergers is $\sim~10^{38}$~erg~s$^{-1}$, so absolute magnitude equals to $\approx -6^\mathrm{m}$. It corresponds to apparent magnitudes $+19^\mathrm{m}$ at the distance 1 Mpc and $+24^\mathrm{m}$ at the distance 10 Mpc. According to \citet{metzger12}, the merger emission's effective temperature is about 5000--7000 K. It corresponds to the LSST g-band. The limiting magnitude in this band is planned to be equal to $+24.8^\mathrm{m}$ for point sources\footnote{\url{https://smtn-002.lsst.io/}} (the limit might become more stringent as the source would be observed with a galaxy as a background). Hence, LSST will be able to detect mergers in galaxies at a few Mpc distances.

The rates above are calculated for the Milky Way, so they might be different for other galaxies. In a particular case of star formation rate independent on time, from eq.~(\ref{eq:sfr_conv}) we obtain:
\begin{equation}
	n(t) = \mathrm{SFR} \times \int\limits_0^t p(\tau)\,d\tau = \mathrm{SFR} \times P(t),
\end{equation}
Here SFR is again in units of stars per year, $P(t)$ is cumulative probability distribution function of mergers in system's age, which is equal to the ratio of the mergers happened until time $t$ to the effective number of systems in the population. This equation is also applicable, if SFR is stable on timescales larger than the timescale of $p(t)$ decay, which is approximately 2-3 Gyr. Let us denote the limit of $P(t)$ at $t \to \infty$ as $P_\infty$. If we again consider only mergers with main sequence stars, for direct impacts $P_\infty = 2.0 \times 10^{-3}$ and for tidal disruptions $P_\infty = 3.2\times 10^{-4}$ (in case of $Q'_{\star} = 10^{5.5}$; for larger $Q'_{\star}$, these values are smaller). Thus for galaxies with constant SFR on large timescales we derive an estimate:
\begin{equation}
	n_\mathrm{di} = 2.0 \times 10^{-3}\, \mathrm{yr}^{-1}\times \mathrm{SFR}
\end{equation}
for direct impacts and
\begin{equation}
	n_\mathrm{td} = 3.2 \times 10^{-4}\, \mathrm{yr}^{-1}\times \mathrm{SFR}
\end{equation}
for tidal disruptions. Here SFR is in units stars per year. The average mass of stars in our population (with Kroupa IMF) is approximately $0.5 M_{\sun}$. Thus, for the Milky Way we obtain $n_\mathrm{di} = 1.2 \times 10^{-2}$ yr$^{-1}$ and $n_\mathrm{td} = 1.9 \times 10^{-3}$ yr$^{-1}$, which are very close to calculated numerically for the same $Q'_{\star} = 10^{5.5}$.

Thus, the higher is SFR -- the more promising is the galaxy for a search of planet-star mergers.

Let us consider also the case of a recent powerful star-formation outburst in a galaxy. There should be a lot of coalescences in it because the probability distribution has maximum at ages about few $\times \, 10^7$ years. E.g., an outburst with SFR=$10\, M_{\sun}\, \mathrm{yr}^{-1}$ and duration of 50 million years creates $\sim 1$~billion stars (depending on the initial mass function, in strong star formation bursts mass distribution can be shifted towards higher masses).
Multiplied by values of $p(t)$ near the maximum, it gives the rate of direct impacts with main sequence stars similar to that in Table~\ref{tab:results} by order of magnitude. On the other hand, there are some obstacles for observing them. In \autoref{sec:results:distr} and \autoref{sec:discus:initial} we show that most of quickly coalescing planets have low masses and massive hosts. For this reason, brightness increase during the event relative to quiescent luminosity of the star can be not sufficiently large for detection. 
Consequently, galaxies with prolonged active star formation are more preferable for searching for mergers.

As for mergers of even very massive planets with red giants or asymptotic giant branch stars, they are more elusive sources \citep{Staff16_giant_interaction}. For this reason we pay less attention to them. Also, our calculations of their rates are less reliable (see \autoref{sec:discus:giants}).

\subsection{Simple estimate of the merger rate}

Numerical results can be compared with an illustrative analytical estimate for the rate of mergers of hot jupiters with their host stars during the main sequence stage.

In the Galaxy there are about $10^{11.5}$ main sequence stars. About 10\% of them belong to FGK spectral classes. About 1\% of FGK stars host hot jupiters with periods <$10$~days \citep{Wright12_occurrence}. Thus, there are $\sim10^{8.5}$ FGK stars with hot jupiters. Planets with orbital periods $\lesssim 3$-6~days can spiral-in due to tides within the lifetime of the Galaxy. If we take the \"Opik law for planetary orbits -- $f(a)\propto a^{-1}$, -- then we roughly obtain $\sim 2\times10^8$ mergers in $10^{10}$~years. I.e., the rate $\sim 0.02$~yr$^{-1}$. This estimate is not far from our results presented in this paper. 
Formally, if in the framework of this simple model we include also red dwarfs, then we obtain the rate an order of magnitude higher, roughly coincident with an estimate by \cite{metzger12}. However, our numerical results indicate that mergers with red dwarf are relatively rare.

A similar estimate can be made for planet consumption by red giants. For the Salpeter mass function stars that can leave the main sequence within the Galactic life time ($M>0.8 M_\odot$) represent about 4-5\%, i.e. $\sim 0.1^{1.35}$, of the whole population of stars with $M>0.08 M_\odot$. This means that $\sim 10^{10}$ of red giants can be formed within $10^{10}$ yrs. Each absorbs few planets. Thus, we obtain the rate few events per year -- in correspondence with the numerical results presented above.

\subsection{Comparison with Previous Studies}

Our results can be compared with estimates made by \cite{metzger12}. These authors obtained the value $\sim 0.1$--1 coalescence between a main sequence host and a hot jupiter per year per Galaxy, assuming $Q'_{\star} = 10^{6}$. If in our numerical calculations we select only mergers of main sequence stars with hot jupiters (using the criterion from \citealt{Wright12_occurrence}), we obtain for the same $Q'_{\star}$ the rate $3\times10^{-3}$~yr$^{-1}$ (for direct impacts, tidal disruptions and stable accretion events altogether). This is about two orders of magnitude lower than the results by \cite{metzger12}.

There are several reasons for this inconsistency. Firstly, our approach to initial conditions and statistics was absolutely different. Instead of using a sample of observed planets biased by selection effects, we used results of population synthesis modeling. Then, we use our population model to calculate the rates of mergers basing on specified stellar initial mass function and star formation rate, i.e., producing a more detailed and accurate normalization than a simpler procedure suggested in \cite{metzger12}.  

Secondly, we used somewhat different formula for tidal evolution of the orbit (compare eq.~(1) in \cite{metzger12} and our eq.~(\ref{eq:adot_beautiful})). The difference is that we took into account stellar rotation. Altogether, the model presented above is more detailed than the approach used by \cite{metzger12}. These authors discuss several simplifications made in their calculations. Here we tried to make a step forward to present a more detailed scenario. Still, in our approach several important simplifications are made (in the first place -- a value of $Q'_\star$ independent on system parameters). Thus, more detailed calculations are welcomed. 

\subsection{Dependence on the Initial Distribution}	\label{sec:discus:initial}

Here we consider the dependence of our results on the initial distribution of planets in mass and initial semi-major axis. Among other things, it gives explanation for unexpected features seen in Fig.~\ref{fig:p_mstar}.

\begin{figure*}
	\centering
	\includegraphics[width=\linewidth]{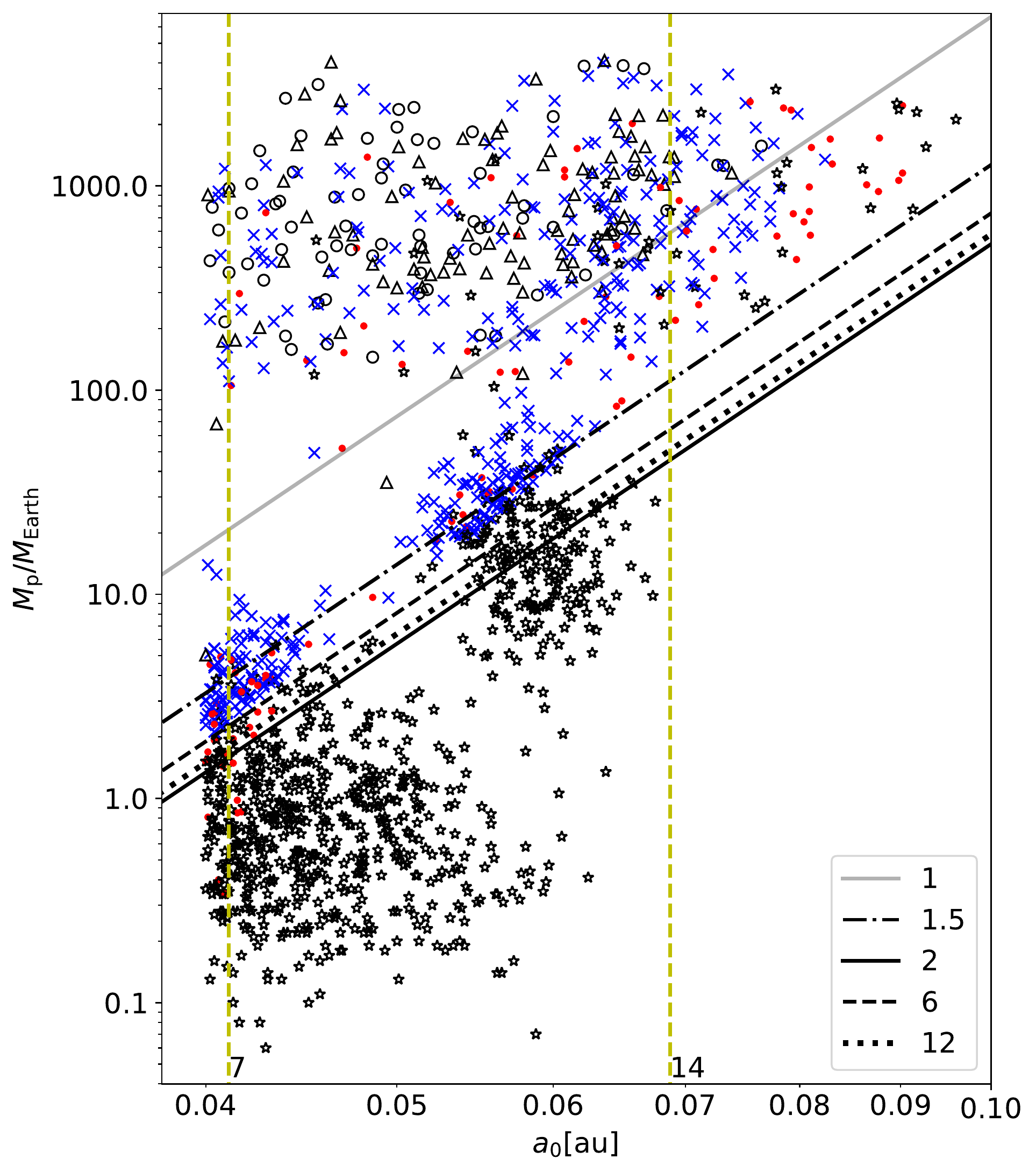}
	\caption{Illustration of the dependence of coalescence statistics on initial distribution in the $a_0-M_\mathrm{p}$ plane. 
    Symbols encode host star masses: circles -- 0.09-1 $M_{\sun}$; triangles -- 1-1.5 $M_{\sun}$; crosses -- 1.5-4 $M_{\sun}$; dots -- 4-7 $M_{\sun}$; stars -- 7-14 $M_{\sun}$. Inclined lines mark the boundary of the ``fall region'' (see text for definition) for stars of different masses (labeled in the legend in units of the Solar mass). The solid line on the top corresponds to 1 $M_\odot$.  Maximum radii of stars of 7 and 14 $M_{\sun}$ at the end of their evolution on the main sequence are plotted by vertical dashed lines and marked by the value of mass in Solar masses.}
	\label{fig:fallenMS_Mstars}
\end{figure*}

To illustrate this dependence graphically, we plot planets that finally undergo coalescence with main sequence hosts during the lifetime of the Galaxy for $Q'_{\star} = 10^{5.5}$ in the $a_0-M_\mathrm{p}$ plane (Fig.~\ref{fig:fallenMS_Mstars}). The total number of planets shown in the plot is chosen by reasons of visual clarity. Different symbols correspond to different stellar mass ranges. 

For a host star with given mass and $Q'_{\star}$ one can define a region in $a_0-M_\mathrm{p}$ space, from which planets are consumed by the star during its main sequence stage (a ``fall region''). If we neglect the change of $R_{\star}$ at the main sequence phase and assume that the star does not rotate, then from the eq.~(\ref{eq:adot_work})
we can express the planetary mass:
\begin{equation} 
	M_\mathrm{p} = \frac{4Q'_{\star}}{117R_{\star}^5}\sqrt{\frac{M_{\star}}{G}}\frac{a_0^{6.5}}{t_\mathrm{fall}},
    \label{eq:maxfall}
\end{equation}
where $t_\mathrm{fall}$ is the time it takes the planet to fall onto the star due to tides. If one substitutes the time since the protoplanetary disk dispersal to the end of stellar main sequence phase in place of $t_\mathrm{fall}$, this equation defines the line in the $a_0-M_\mathrm{p}$ plane, above which all planets that fall onto the star during its main sequence phase are situated\footnote{Note that the dots in Fig.~\ref{fig:fallenMS_Mstars} are calculated numerically using the model including stellar rotation, while the lines are drawn for non-rotating stars. If we neglect stellar rotation also for dots, they would lie closer to the corresponding lines, but still above them.}. In Fig.~\ref{fig:fallenMS_Mstars} a set of such lines is plotted. The values of stellar mass for them are 1, 1.5, 2, 6, and 12 $M_{\sun}$. 
Values of stellar radius in eq.~(\ref{eq:maxfall}) correspond to an intermediate value during the main sequence stage for each mass: $\log R_\star=0.5(\log R_\star(t_0)+\log R_\star(t_\mathrm{f}))$, where $t_0$ and $t_\mathrm{f}$ are the first and the final points of the track for the main sequence stage. 

It is clearly seen that the line shift with changing stellar mass is not monotonous. For $M_{\star} \lesssim 1.5M_{\sun}$, only small part of planets (almost only hot jupiters) lies above the line (in the ``fall region''). Then, if we increase the mass up to $\sim2 \, M_\odot$ such a line quickly moves down, i.e. more and more planets appear in the ``fall region''. If $M_{\star} \approx 2M_{\sun}$, then the ``fall region'' includes the largest number of planets, which causes the first peak in Fig.~\ref{fig:p_mstar}. With further increasing of the stellar mass, the line shifts back (i.e., goes up in the plot). It corresponds to  low probability of a coalescence at $4M_{\sun} \lesssim M_{\star} \lesssim 7M_{\sun}$ seen in Fig.~\ref{fig:p_mstar}. The second peak in Fig.~\ref{fig:p_mstar} is due to two reasons. Firstly, for $M_{\star} \gtrsim 7M_{\sun}$ the boundary of the ``fall region'' shifts downward again. Secondly, high-mass stars expand significantly during their main sequence phase (according to the evolutionary tracks we use), and large number of planets lying outside of the ``fall region'' are engulfed by the star not by tides, but in the process of expansion. To illustrate it, for stellar masses of 7$M_{\sun}$ and 14$M_{\sun}$ in Fig.~\ref{fig:fallenMS_Mstars} we plot the stellar radius at the end of the main sequence stage by vertical dashed lines. One can see that the fallen planets outside of the ``fall region'' lie inside the maximum radius of their host stars. (Several dots apparently violate this rule. However, this is because ``fall region'' boundaries on the plot are only approximate due to simplified averaging of stellar radius in time.) 

This is the evidence that the numerous mergers of Earth and Neptune mass planets with massive stars ($\sim 10M_{\sun}$), occurring in our simulation during the first few tens million years after system formation, have non-tidal nature and caused by the expansion of the star at the main sequence stage.

The considered examples show that details of the initial distribution dramatically affect the rate at which a star of a given mass swallows planets. That means, $a_0-M_\mathrm{p}$ distribution of planets is a very important ingredient of the model and setting an incorrect distribution leads to serious uncertainties which can crucially influence final conclusions. 

In our opinion, the main source of uncertainties of this kind in our calculations is due to the fact that the distribution we use has the same form for all stars, while only the number of planets per system is varying with stellar mass (see e.g. \citealt{2018haex.bookE.153M}). The form of the assumed distribution follows the results of \citet{Alibert13} for planets around Solar-mass stars. Meanwhile, both observations (e.~g.~\citealt{Gillon17_trappist1}) and simulations (e.~g.~\citealt{Alibert17_lowmass}) show that low-mass stars have more compact planetary systems than FGK-stars (for example, due to smaller masses of protostellar nebula and due to smaller distance to the snow line). Probably, it makes possible a significant number of planet-star coalescences also in M-dwarf systems. Some studies (e.~g.~\citealt{Alibert11}) also predict that massive stars have less close-in planets than Solar-mass ones, so fraction of mergers with massive stars in the total rate may be not as significant as in our model. Furthermore, our eq.~(\ref{eq:n_pl}) for the number of planets per system is based on just qualitative consideration. Therefore, accurate treatment of distribution changing with stellar mass is needed.

In addition, in the calculations presented above we based all our assumptions about the $a_0-M_\mathrm{p}$ distribution on a particular model of formation and early evolution. Intensive planet-planet interaction during early period of life of a planetary system can significantly modify the shape of $a-M_\mathrm{p}$ plot respect to the one used in our modeling, as well as the eccentricity distribution (see, e.g. \citealt{2008ApJ...686..603J, 2011Natur.473..187N, 2011ApJ...735..109W, 2016MNRAS.456.3671A}). It might be very difficult to obtain the detailed distribution purely from observations in the near future, thus most probably future, more detailed, population synthesis calculations of young planetary systems can provide necessary data.

\subsection{Tidal Model Uncertainties}

To calculate tidal evolution of planetary systems, we use eq.~(\ref{eq:adot_work}) with the same (in a given set of calculations) constant $Q'_{\star}$ for all main sequence star systems. This approach is widespread \citep{Dobbs-Dixon04_tides, Jackson08, metzger12, Penev12_Q, Penev14_POET, Zhang_dizzy,Ginzburg17_coremass}. However, the values of $Q'_{\star}$ adopted by these authors are different by orders of magnitude. As can be seen from the presented results, variation of this parameter significantly modifies the coalescence rate.

The model with fixed $Q'_{\star}$ is oversimplified as $Q'_{\star}$ might be different for stars with different masses and vary with planet's period and mass. It is also different for a given star at different ages. For example, \citet{2017A&A...604A.112G} and \citet{2019A&A...628A..42H} calculated that for solar-mass stars tidal quality factor corresponding to the dissipation of dynamical tide in the convective envelope (see below) evolves from $10^4$ in the beginning of MS phase up to about $10^9$ in its end.
In fact, usage of $Q'_{\star}$ for tides on stars is just a parametrization of complex tidal processes by the single parameter. There are several different approaches developed for modeling tidal dissipation in stars. Below we briefly outline them.

In the tidal theory response of a host star on  perturbations from its planet can be divided into two components: equilibrium and dynamical tides. The first refers to the tidal bulge formation, while the second -- to the oscillations of the star forced by the tidal potential.

\citet{Zahn77,Zahn89} derived an equation for orbital evolution driven by equilibrium tide dissipation in stellar convective envelope due to turbulent viscosity. Zahn did not use Q-parametrization, and his equation has a different form in comparison with eqs.~(\ref{eq:adot_beautiful}), (\ref{eq:adot_work}). A more useful equation based on his theory is given by \citet{Privitera16_interactionsI}. \citet{Penev_Sasselov} performed numerical simulations of equilibrium tide dissipation in convective envelopes and obtained $Q'_{\star} \gtrsim 10^8$. It corresponds to orders of magnitude lower dissipation than derived by \citet{Jackson08} from eccentricity distribution of exoplanets.

There are also many studies on dynamical tides (see, e.g. \citealt{2017MNRAS.470.2054C} and references therein). The main channels of their dissipation are inertial waves damping in the convective envelopes and internal gravity waves (g-modes) damping in radiative regions. \citet{2018A&A...618A..18R} found that the dissipation of inertial waves in convective envelopes of Solar-mass stars does not have significant effect on planetary orbits during the main sequence and post-main sequence stages. In the above-mentioned paper by \citet{Zahn77} damping of g-modes in stars with convective core and radiative envelope has been considered. However, only a small part of planets orbits such stars. More recent studies concern g-modes damping in radiative cores of Solar-type stars. The primary waves in such stars are excited near the core-envelope boundary. As they propagate inwards, their amplitude increases due to geometrical focusing, and non-linear effects become significant. If the mass of close-in planet is more than about 3 Jupiter mass, waves overturn and break in the center of the star (the strongly non-linear regime; see \citealt{Barker_Ogilvie_10, Barker11, Barker_Ogilvie_11}). 
If the planet's mass lies in range from 0.5 to 3 Jupiter mass, then tide evolves in the weakly non-linear regime. Primary waves do not break, but excite lots of secondary waves of shorter wavelengths. These secondary waves can dissipate effectively \citep{Essick_Weinberg16}. The calculations in the works mentioned above show that dynamical tides provide several orders of magnitude higher dissipation rate for Solar-mass stars than equilibrium tides ($Q'_{\star}$ down to $\sim10^5$). As well, they show strong dependencies between $Q'_{\star}$ and planetary mass and period. However, the limitations of their applicability are very strict. For example, \citet{Essick_Weinberg16} present the fit of their numerical calculations $Q'_{\star} \propto M_\mathrm{p}^{0.5} P^{2.4}$ (where $M_\mathrm{p}$ is the planet mass and $P$ is the orbital period) for $P\lesssim 4$~days for solar-mass star, which corresponds to $a\lesssim 0.05$~au. For wider orbits they show that non-linear effects are weaker and the dependence is less strong. Since the inner boundary of initial semi-major axis distribution in our simulation is 0.04~au, the regime studied by \citet{Essick_Weinberg16} for most planets plays a role only during the final stages of inspiral, which are very short (see Figs.~\ref{fig:single_change_a0}-\ref{fig:single_change_ms}).

Summarizing the discussion above, we note that in our study we explore a reasonable range of $Q'_{\star}$ values, but we neglect its dependence on system parameters, which leads to uncertainties and potentially can modify our results.

\subsection{Mergers with post-main sequence stars}  \label{sec:discus:giants}

There are some issues in our calculation of the rate of coalescence of planets with post-main sequence stars that should be commented on. Firstly, we use only one value -- $Q'_{\star}=10^{5.5}$, -- for all post-main sequence stars. Due to large differences between normal and post-main sequence stars in terms of internal structure, in reality $Q'_{\star}$ at later stages of stellar evolution might differ significantly from the value for main sequence stars,
and also can be different for different giant stars (depending on their masses, exact stage of evolution, etc.). At the same time, as was already mentioned, \cite{2017ApJ...849L..11W} obtained the value $Q'_\star = 2 \times 10^5$ for the subgiant WASP-12, which is close to the value $Q'_{\star}=10^{5.5}$ used in our model.

Secondly, we set the outer boundary of planetary initial semi-major axis distribution to 2 au, because most of sufficiently massive planets in the original distribution of \citet{Alibert13} lie within this radius. However, there are planets beyond this radius, too. Their orbits do not undergo considerable tidal evolution, but they can be swallowed by expanding post-main sequence stars. According to evolutionary tracks, the most massive stars in our simulation ($14 M_{\sun}$) have at the end of their evolution radii close to 5 au. Consequently, we underestimate the rate of coalescence of planets with post-main sequence stars (for the given $Q'_{\star}$). Still, the fraction of stars that expand up to such large radii (i.e., those that are massive enough) is very low, thus underestimation seems to be low, too.

Also it should be mentioned that in the mass range of stars from 0.75 to 1.95 $M_{\sun}$ our code does not follow stellar evolution till its end. For stars within this range we do not treat the horizontal branch stage. According to the evolutionary tracks by  \cite{Bressan12_tracks},  after the horizontal branch stage a star expands more than at the red giant stage only for masses approximately from 1.5 to 2 $M_{\sun}$. For this reason we do not expect that inclusion of the horizontal branch evolution of all stars in our treatment would have significant effect on the rate of coalescences.

Finally, the luminosities of mergers of planets with post-MS stars presented in our work are rather upper estimates than exact calculations. As was described in \autoref{sec:model:lum}, for such events we take the minimum of the values given by eqs.~(\ref{eq:l_di}) and (\ref{eq:l_giant}). The former was derived by \citet{metzger12} for the case of a main sequence star under the assumption that the optical emission from the merger is formed by hydrogen recombination in the outflows caused by the interaction of planet with the outer layers of the star. The latter simply equals the luminosity to the orbital energy decay rate. The applicability of both approaches for merger of a planet with a red giant is a disputable question. That is why the significant rate of bright transients from mergers of planets with post-main sequence stars (about $10^{-2}~\mathrm{yr}^{-1}$) is just an upper estimate.

\subsection{Other Sources of Uncertainties}
\label{sec:discus:other}

Obviously, there can be many other sources of uncertainties in our population synthesis scenario. Below we briefly comment on few of them.

In our calculations we assumed zero eccentricity. This is, of course, a significant simplification. Tidal evolution can be much faster for non-zero eccentricities, especially for terrestrial planets, see e.g. \cite{2004ApJ...614..955M}. Non-zero eccentricities can be easily obtained by planets on early stages of evolution. However, as here we are mostly interested in calculations of the coalescence rate of massive planets, our assumption is at least partly justified: for jupiter-like objects the effect of non-zero eccentricities is smaller, and being massive they typically attain lower eccentricities than, for example, earth-like planets.

In this study we used evolutionary tracks just for one value of the metallicity: $Z=0.02$. No doubts, due to several reasons the rate of planet-star coalescence might depend on it (see \citealt{2018haex.bookE.153M} and references therein). In the first place, the following planetary properties might depend on the metallicity: number of planets in a system, their mass (and, probably, initial semi-major axis) distribution. Then, stellar properties depend on $Z$ so that we can expect different values of $Q'_\star$, different rotation rates, and, more important, different stellar evolution. Even for a given galaxy stars have different metallicities (especially those, formed at early years of the galaxy). However, we demonstrated that for estimates of the rate of planet-star mergers mostly relatively recent (few Gyrs) star formation is important. Still, galaxies can have different metallicities on the whole (a simple example -- Magellanic clouds in comparison with the Milky Way). On the other hand, most  massive galaxies with significant star formation within last few Gyrs at distances below $\lesssim 10$~Mpc have approximately solar metallicity. Anyway, some differences can be expected. And of course, our results might be significantly modified to apply them to elliptical galaxies. 

Even if distributions of several parameters are well-known, there can be an important source of uncertainties related to correlations between different distributions. In our model we assumed that many parameters are not correlated with each other. For example, we assumed that the crucial ingredient of the considered scenario -- initial $a-M_\mathrm{p}$  distribution, -- is the same for all masses of host stars. Most probably, this is not the case in reality (see \autoref{sec:discus:initial}).
Unfortunately, at the moment we are not aware of detailed modeling of this effect which can allow to produce a realistic set of initial conditions depending on the mass of a host star.
Another possible correlation can appear due to formation of planets around stars in dense groups in the case of high SFR. Thus, distribution of planetary orbits and lifetime of protoplanetary discs (as well as several other parameters) can be correlated with the SFR. 

Note, that the SFR by itself is an important parameter and it influences the final results. In this paper we used the rate defined in eq.~(\ref{eq:sfr_steps}). Even for the Milky Way this quantity is not well known. However, in the case of our Galaxy it introduces uncertainty less than a factor of two. 

Another source of uncertainty is rough calculation of $a_\mathrm{sync}$ (see \autoref{sec:model:tidal}) for stars with mass larger than $1.25M_{\sun}$. In this case, we assume invariability of $a_\mathrm{sync}$ in time and its single-valued dependence on stellar mass, as well as uniform rotation and homogeneous structure of stars. However, the resulting $a_\mathrm{sync}$ for stars slightly more massive than $1.25M_{\sun}$ corresponds to $a_\mathrm{sync}$ for the stars slightly low massive than this threshold at the end of the MS phase. Since for low-mass stars rotational evolution was calculated more accurately, this correspondence indicates that the estimate for more massive stars is reasonable.

We made our calculations not only for the model of planet orbital evolution described in \autoref{sec:model:tidal}, but also with the model  neglecting stellar rotation (i.~e., without the factor $\left[1 - \left(P_\mathrm{orb}/P_\mathrm{rot}\right)\right]$ in eq.~(\ref{eq:adot_beautiful}) and without any rotational evolution). In this simplified model, merger rates are up to 2 times higher than in the model presented above. The influence of the introduction of stellar rotation into the model is different for different types of mergers and for different $Q'_{\star}$. It is insignificant for the statistics of planet engulfments by post-MS stars. As for mergers with host stars at the MS, the effect is maximal for direct impacts in case of small $Q'_{\star}$ and for tidal disruptions. We also found that, unlike the change of the equation of tidal orbital evolution, accounting for rotational evolution of low-mass and solar-mass stars practically does not change the merger statistics, because practically all mergers occur with stars more massive than $1.5M_{\sun}$.

Lastly, with regard to other sources of uncertainties, we 
do not take into account influence of binary companions, use approximate equation for luminosities (especially in case of tidal disruptions and for mergers with post-MS stars). Of course, this might influence the final results, but we hope that our estimates can still be valid by an order of magnitude.

\section{Conclusions}	\label{sec:concl}

We presented a population synthesis model to calculate the rate of star-planet mergers. For initial conditions we used results of population synthesis studies by \cite{Alibert13}. We calculate the tidal evolution of planetary orbits following the equation from \cite{2004ApJ...614..955M} for a set of tidal quality factors $Q'_{\star}$. In our calculations of the coalescence output we follow \cite{metzger12}. 

We obtain that the coalescence rate is dominated by consumption of planets by red giants with the rate $\sim 3$~yr$^{-1}$ per Milky Way-like galaxy. However, such events very rarely produce observable transients. 

Mergers with main sequence stars are dominated by low-mass planets merging with stars $M\gtrsim1.5 M_\odot$. Their rate is $\sim 0.006-0.009$~yr$^{-1}$ per galaxy (depending on $Q'_{\star}$). These type of coalescence produce transients with $L\sim 10^{35} - 10^{37}$~erg~s$^{-1}$ -- not bright enough to be easily detectable from Mpc distances. 

Brighter events are due to mergers of massive planets with main sequence stars. In the simulation with most efficient tidal dissipation in our set, they happen with the rate $\sim 0.003$~yr$^{-1}$. Having luminosities $\sim10^{37.5}$~--~$10^{38}$~erg~s$^{-1}$, they can be observed from nearby galaxies up to 10 Mpc distance with the next generation of optical surveys.  These rates are significantly lower than those presented by \cite{metzger12}. We attribute it to different initial distributions and different approach to calculate the final normalization for the Galactic rate.

Our model contains significant simplifications related to parameters of the tidal evolution. Detailed calculations based on more reliable initial distributions (obtained from observations or more sophisticated modeling) and more accurate treatment of tidal dissipation are welcomed considering future observations with LSST.

\section*{Acknowledgements}

We thank the anonymous referee of the first variant of this paper for useful comments and suggestions and Sergei Chernov for a fruitful discussion. SBP acknowledges support from the Program of development of M.V. Lomonosov Moscow State University (Leading Scientific School ``Physics of stars, relativistic objects and galaxies''). AVP was supported by the government of the Russian Federation (agreement 05.Y09.21.0018).




\bibliographystyle{mnras}
\bibliography{exoplanets}

\bsp	
\label{lastpage}
\end{document}